\documentclass[
  aip,
  amsmath,
  amssymb,
 reprint,
 onecolumn
]{revtex4-1}

\usepackage{graphicx}
\usepackage{dcolumn}
\usepackage{bm}

\usepackage{lineno}
\usepackage{lineno}
\usepackage[utf8]{inputenc}
\usepackage[T1]{fontenc}
\usepackage{mathptmx}
\usepackage{etoolbox}

\usepackage{comment}
\usepackage[dvipsnames]{xcolor} 
\usepackage[hidelinks]{hyperref} 
\newcommand{\blue}{\color{black}} 

\makeatletter
\def\@email#1#2{%
 \endgroup
 \patchcmd{\titleblock@produce}
  {\frontmatter@RRAPformat}
  {\frontmatter@RRAPformat{\produce@RRAP{*#1\href{mailto:#2}{#2}}}\frontmatter@RRAPformat}
  {}{}
}%
\makeatother

\begin{document}


\title{Extension of the law of the wall exploiting weak {\blue similarity} of velocity fluctuations in {\blue turbulent channels}}

\author{Christoffer Hansen}
\affiliation{Department of Mechanical and Production Engineering, Aarhus University, 8200 Aarhus N, Denmark}

\author{Jens N. Sørensen}
\affiliation{Department of Wind and Energy Systems, Technical University of Denmark, 2800 Lyngby, Denmark}

\author{Xiang I. A. Yang}
\affiliation{Department of Mechanical Engineering, Pennsylvania State University, State College, PA, 16802, USA}

\author{Mahdi Abkar}
\affiliation{Department of Mechanical and Production Engineering, Aarhus University, 8200 Aarhus N, Denmark}

\email{xzy48@psu.edu, abkar@mpe.au.dk}

\date{\today}
            
\begin{abstract}
This paper explores the {\blue similarity} of the streamwise velocity fluctuations in a channel.
In the analysis, we employ a one-dimensional scalar variant of the proper orthogonal decomposition (POD).
This approach naturally motivates the introduction of two different levels of {\blue similarity} which we will refer to as strong and weak {\blue similarity}.
Strong {\blue similarity} requires that {\blue the two-point correlation, and thus, all POD modes, show Reynolds number similarity}, while weak {\blue similarity} only requires that the first few POD modes {\blue show similarity}.
As POD concerns information at more than one location, these {\blue similarities} are more general than various similarities found in the literature concerning single-point flow statistics.
We examine flows at $Re_\tau=180$, 540, 1000, and 5200.
Strong {\blue similarity} is observed in the viscous layer and the wake region, and weak {\blue similarity} is found in both the viscous wall region and the outer part of the logarithmic layer.
The presence of weak {\blue similarity} suggests the existence of an extension to the law of the wall (LoW).
We propose such an extension based on the results from the one-dimensional POD analysis.
The usefulness of the LoW extension is then assessed by comparing flow reconstructions according to the conventional equilibrium LoW and the extended LoW.
We show that the extended LoW provides accurate flow reconstructions in the wall layer, capturing fine-scale motions that are entirely missed by the equilibrium LoW. 
\end{abstract}

\maketitle


\section{Introduction}
\label{sec:introduction}

The mean velocity profile in an incompressible equilibrium boundary layer is described by the law of the wall (LoW).
With $u_\tau = \sqrt{\tau_w / \rho}$ denoting the mean friction velocity, where $\tau_w$ is the mean wall-shear stress and $\rho$ is the density of the fluid, the LoW states that the wall-unit-scaled mean velocity $U^+=U/ u_\tau$ is universal in the logarithmic layer.
Specifically, $U^+$ is linearly proportional to the logarithm of the wall-unit-scaled distance from the wall, i.e., $U^+ = \ln(y^+)/k + B$, where $y^+ = (u_\tau / \nu)y$, $\nu$ is the kinematic viscosity, $k$ is the von Kármán constant, and $B$ in another constant.{\blue \cite{Prandtl_1925_log_law,von_Karman_1930_Log_law}}
{\blue The LoW is supported by high Reynolds numbers laboratory experiments, i.e., for $Re_\tau \gtrapprox 10.000$, where $Re_\tau = u_\tau \delta / \nu$ is the friction Reynolds number.\cite{Smits_et_al_2011_High_Re_wall_turbulence_2,marusic2013logarithmic}
Here, $\delta$ is the half-channel height, pipe radius, or momentum thickness depending on whether channel flows, pipe flows, or boundary layers are considered, respectively.
At lower Reynolds numbers, results from direct numerical simulations (DNS) show that the mean velocity profile does not strictly follow the LoW, but that the LoW nonetheless remains a good approximation.\cite{Sillero_et_al_2011_DNS_BL_6650,Lee_Moser_2015_DNS_chan_5200,Pirozzoli_et_al_2021_DNS_pipe_6000}
The LoW has therefore long served as an anchor point for turbulence modeling.\cite{spalart1992one, kalitzin2005near, bin2022progressive}}

An important application of the LoW is in the context of large-eddy simulation (LES) wall modeling.{\blue \cite{piomelli2002wall,Bose_et_al_2018_WMLES_review}}
In equilibrium wall models, the LoW is usually invoked for both of the wall-parallel velocities.{\blue \cite{kawai2012wall,yang2017log}}
However, in this work, we choose to focus on the streamwise velocity component as it is generally more important than its spanwise counterpart.
The approximation of the streamwise velocity in equilibrium wall models can be written generically as
\begin{equation}
    u({\bf x},t) \approx c(x,z,t){\rm LoW}(y^+)  \, ,
    \label{eq:LoW}
\end{equation}
where ${\bf x}$ denotes a given location in the flow, $t$ is the time, and $c$ is a local coefficient.
We will refer to $x$, $y$, and $z$ as the streamwise, wall-normal, and spanwise coordinates, respectively.
Further, note that we take ${\rm LoW}$ to be nondimensional such that the friction velocity $u_\tau$ has been absorbed into the coefficient $c$.
While many have obtained acceptable results for mean velocity profiles using equilibrium wall models in the context of LES,{\blue \cite{lehmkuhl2016flow,lehmkuhl2018large,goc2022large}} the practice of imposing the LoW locally and instantaneously has received much criticism.
{\blue Several authors} have argued that while the mean flow follows the LoW in a 2D turbulent channel, the velocity fluctuations do not.{\blue {\cite{park2014improved,yang2015integral,lv2021wall}}
}
Similarly, in flows with non-equilibrium effects, the mean flow will no longer follow the LoW.
This suggests that the LoW must be augmented to account for flow unsteadiness and other non-equilibrium effects.{\blue \cite{Bose_et_al_2018_WMLES_review}} 

The idea of extending the LoW for the purpose of wall modeling in LES has been pursued by a number of authors.{\blue \cite{yang2015integral,lv2021wall}}
{\blue These authors} analyzed the streamwise Reynolds-averaged momentum equation and argued that a linear term $y^+$ should be the leading-order extension to the LoW.
Specifically, they proposed the following approximation
\begin{equation}
   u({\bf x},t) \approx c(x,z,t){\rm LoW}(y^+) + a(x,z,t)y^+ \, , 
   \label{eq:lineary}
\end{equation}
where $c$ and $a$ are both coefficients.
There, \eqref{eq:lineary} was interpreted as a first-order truncation of the Taylor expansion of the velocity profile in the vicinity of the LoW
\begin{equation}
    u({\bf x},t) \approx c(x,z,t){\rm LoW}(y^+) + \sum_{j} a_j(x,z,t) (y^+)^j \, . 
   \label{eq:poly}
\end{equation}
Including more terms on the right-hand side naturally provides a more accurate approximation.
However, more information is also required to determine the coefficients in front of these terms. 
In the context of LES wall modeling, this could be accomplished, e.g., by using a matching condition with the LES alongside physics-based constraints or by increasing the number of matching conditions.{\blue \cite{yang2015integral,hansen_yang_abkar_2023}}
Nonetheless, from the perspective of computational efficiency and code usage, it is preferred to have as few terms as possible in the expansion.
Therefore, the question is if \eqref{eq:lineary}, or similar variants using generic basis functions, is a good starting point for velocity reconstruction.
In this regard, we argue that it is unlikely that a few polynomials or other mathematical basis functions can efficiently capture turbulent motions.
Thus, we ask if a set of basis functions can be determined, most importantly, a set of universal basis functions, that can optimally capture the flow in the wall layer.

The above discussion strongly motivates the application of modal analysis,{\blue \cite{taira2017modal}} which could allow for the extraction of a more physical and efficient extension to the LoW.
Applying modal analysis, we may write the streamwise velocity as a linear superposition of the LoW and a series of modes
\begin{equation}
     u ({\bf x}, t) \approx c(x,z,t){\rm LoW}(y^+) + \sum_{j} a_j\varphi_j \, ,
     \label{eq:general}
\end{equation}
where $\varphi_j$ are the modes and $a_j$ are the modal coefficients.
Note that the temporal and spatial dependence of $\varphi_j$ and $c_j$ are left out deliberately at this stage.
In this work, we will consider proper orthogonal decomposition (POD) analysis.{\blue \cite{berkooz1993proper,holmes2012turbulence}}
In most cases, when applying POD to a 3D scalar field, e.g., the streamwise velocity component, the expansion in \eqref{eq:general} is written as
\begin{equation}
    u({\bf x},t) \approx c(x,z,t){\rm LoW}(y^+) +\sum_{j=1}^n a_j(t)\varphi_j({\bf x}) \, ,
    \label{eq:POD}
\end{equation}
where $n$ is the number of POD modes included in the expansion.

Although modal analysis is well developed, it has not yet had an impact on predictive modeling in LES wall modeling.
We believe that this is largely because of the following three major obstacles.
First, most modal analyses are performed in 2D or 3D space; 2D or 3D modes will inevitably be strongly geometry-specific, and thus, {\blue cannot be universal over a range of different flows}.
Second, structures with an increasingly large streamwise extend are known to appear in the log-layer as the Reynolds number increases\cite{mathis2009large,marusic2010wall}; 2D or 3D modes are therefore unlikely to {\blue show Reynolds number similarity in a given geometry}.
Third, 2D or 3D modes are non-local in the wall-parallel directions; this greatly complicates the coupling between the resulting wall model and the LES solver.
The third obstacle can be overcome by rewriting \eqref{eq:general} to mimic \eqref{eq:poly} as follows
\begin{equation}
    u({\bf x},t) \approx c(x,z,t){\rm LoW}(y^+) + \sum_{j=1}^n a_j(x,z,t) \varphi_j(y^+) \, ,
    \label{eq:PODy}
\end{equation}
such that the modal decomposition is limited to be along the wall-normal direction.
Here, $\varphi_j$ are one-dimensional POD modes and $a_j$ are coefficients depending on both space and time.
Since the modes in \eqref{eq:PODy} are local, applying this approach for wall modeling is equally straightforward as applying \eqref{eq:poly}, which solves the third obstacle.
Further, regarding the second obstacle, the appearance of turbulent structures with an increasingly large streamwise extent that grows with the Reynolds numbers should now correspond to a change of the coefficients in \eqref{eq:PODy} instead of a change in the modes themselves.
{\blue However, even though one-dimensional modes are inherently less Reynolds number dependent and geometry-specific than 2D or 3D modes, they are still not guaranteed to show Reynolds number similarity or to be universal between different flows.
The topic of the present work is to address the second obstacle, i.e., to investigate the existence of an extension of the LoW, based on one-dimensional POD modes in the form of \eqref{eq:PODy}, which shows Reynolds number similarity in turbulent channel flows.
The possibility of overcoming the first obstacle, which would require POD modes that are universal over a range of different flows, will be considered in future works.}

To that end, we will define two levels of {\blue similarity}: strong {\blue similarity} and weak {\blue similarity}.
The existence of strong {\blue similarity requires that the two-point correlation itself shows Reynolds number similarity}, which would imply that the modes $\varphi_j$ {\blue show similarity} with respect to the Reynolds number for all $j$. 
The existence of weak {\blue similarity} would, on the other hand, imply that the modes $\varphi_j$ only {\blue show similarity} for $j=1,2,\ldots,N$ for some finite $N$.
As \eqref{eq:PODy} is an expansion of the velocity profile around the LoW, the modes $\varphi_j$ can be interpreted as extensions of the LoW.
An extension {\blue which shows Reynolds number similarity} would minimally require that it holds for one mode, $\varphi_1$, which corresponds to the existence of weak {\blue similarity} for $N=1$.

In anticipation of the discussion in the next sections, we note that we are not the first to study the {\blue similarity} of POD modes in wall-bounded turbulent flows.
For pipes, self-similarity of the wall-normal part of two-dimensional POD modes (wall-normal and azimuthal directions) {\blue has been observed.\cite{hellstrom2016self}}
Specifically, these authors observed that the wall-normal part of the POD modes could be collapsed across different azimuthal wavenumbers using a scaling of the wall-normal coordinate based on the peak position of the POD modes.
We note that this differs from weak {\blue similarity} as defined in this work which focuses on the {\blue similarity} of POD modes across different Reynolds numbers rather than the self-similarity of different POD modes at the same Reynolds number.
For channels, weak {\blue similarity} of the one-dimensional streamwise POD modes along the wall-normal direction in the wake region of a turbulent channel {\blue has previously been reported.\cite{liu1994reynolds}}
Similarly, weak {\blue similarity} of two-dimensional POD modes over different types of rough walls in the outer part of the flow {\blue has also been observed.\cite{placidi2018turbulent}}
Finally, {\blue some authors have} tried to derive analytical expressions for the one-dimensional POD modes of the streamwise velocity over the full wall-normal extend of a turbulent channel.{\blue \cite{carbone1996hierarchical,baltzer2011structure}}
However, contrary to the current study, these authors tried to capture the change in the POD modes with Reynolds number rather than looking for modes {\blue which show Reynolds number similarity.}

The remaining parts of this work are organized as follows.
In \S\ref{sec:POD_section}, we present details of the one-dimensional POD analysis and elaborate on the concepts of strong and weak {\blue similarity}.
In \S\ref{sec:results}, we present the results regarding the existence of strong and weak {\blue similarity}.
An extension of the LoW is then proposed based on the observed weak {\blue similarity} and used for flow reconstruction to test its descriptive power. 
Finally, in \S\ref{sec:conclusion}, we present some concluding remarks.
\section{Methodology}
\label{sec:POD_section}
We give an overview of POD and discuss its one-dimensional scalar variant in \S\ref{subsec:General_background}.
We then define two levels of Reynolds number {\blue similarity} in \S\ref{subsec:similarity_in_POD}.
We will make a connection between these two levels of {\blue similarity} and the {\blue similarity} of the modes $\varphi_j$ in \eqref{eq:PODy}.
Finally, the details of the DNS channel flow datasets used in this work are summarized in \S\ref{subsec:DNS_channel_datasets}.

\subsection{POD}
\label{subsec:General_background}
The POD was originally introduced within the fluid mechanics and turbulence communities by Lumley.{\blue \cite{Lumley_POD_paper_1967,Lumley_Stochastic_tools_1970}}
Since then, POD has become a widely used tool for identifying coherent structures in fluid flows,{\blue \cite{Moin_et_al_POD_channel_flow_1989,berkooz1993proper,taira2017modal}} as well as a central component of dynamical system models of fluid systems.{\blue \cite{aubry1988dynamics,Noack_et_al_POD_Galerkin_Cylinder_2003,holmes2012turbulence,rowley2017model}}
Recently, several authors have reviewed the details of POD to foster a greater level of clarity regarding the different variants of the method.{\blue \cite{George_POD_review_2017,Towne_et_al_SPOD_2018}}
In the majority of existing literature, POD is performed in three-dimensional space using a vector variant that covers all three components of the fluctuating velocity.
This makes sense from both the coherent structure and dynamical system modeling perspectives for the following reasons.
In turbulent flows, coherent structures are inherently three-dimensional quantities and involve specific relations between the different velocity components, making the three-dimensional vector variant of POD the obvious choice.
In fact, it has been argued that coherent structures are inherently $(3,1)$-dimensional (3 spatial and 1 time), and thus, a space-time vector variant of POD should provide structures that more closely mimic those found in real fluid flows.{\blue \cite{Towne_et_al_SPOD_2018}}
For dynamical system modeling, POD is typically used to provide a linear subspace onto which the Navier-Stokes equations can be projected using the Galerkin method.{\blue \cite{holmes2012turbulence}}
In this case, the 3-dimensional vector variant of POD delivers the most compact dynamical system model in terms of the number of degrees of freedom, while a scalar variant would result in a more complex dynamical system model with a higher number of degrees of freedom.{\blue \cite{Rowley_et_al_POD_Galerkin_comp_fluid_2004}}

In this work, we resort to a one-dimensional scalar variant of POD for the following reasons. 
Firstly, the purpose of this work is not to identify coherent structures or construct dynamical system models; we instead aim to investigate the existence of an extension of the LoW.
This justifies the choice of a 1D variant of POD as discussed in \S\ref{sec:introduction} above.
Secondly, since we focus on the streamwise velocity component in this work, it makes sense to employ a scalar variant of POD.

We present the details of the one-dimensional scalar POD analysis here.
Consider a real-valued zero-mean one-dimensional scalar stochastic process $\{ q(y,Re_\tau) \}$ where $y$ is the only independent variable and the friction Reynolds number $Re_\tau = u_\tau \delta / \nu$, with $\delta$ the channel half-width, represents the only parameter dependence.
Note that the Reynolds number dependence is suppressed in the presentation below for notational clarity.
Such a stochastic process can be defined by sampling realizations of the streamwise velocity fluctuations $u'(y)$ along the wall-parallel directions $x$ and $z$ as well as in time $t$.
We also  define an inner product
\begin{equation}
    (f,g)_\Omega = \int_\Omega f g \, dy \, ,
    \label{eq:inner_prod}
\end{equation}
where $\Omega$ is the wall-normal interval under consideration and both $f$ and $g$ are real-valued functions.
Finally, we define an averaging operation $\langle \, \cdot \, \rangle$ where the averaging is performed over $x$, $z$, and $t$.

POD can now be regarded as an optimization problem where the goal is to find the deterministic modes $\varphi(y)$ which maximize the normalized ensemble-averaged norm-squared projection
\begin{equation}
    \lambda = \frac{\langle | \left( q(y) , \varphi(y) \right)_\Omega |^2 \rangle}{( \varphi,\varphi)_\Omega} \, .
\end{equation}
Rewriting this problem using the calculus of variations leads to the following Fredholm eigenvalue problem
\begin{equation}
\label{eq:fredholm_eig_problem}
    \int_\Omega C(y,y')  \varphi(y') \, dy' = \lambda \varphi(y) \, ,
\end{equation}
where the two-point correlation for the stochastic process is given by
\begin{equation}
    C(y,y') = \langle q(y) q(y') \rangle \, .  
\end{equation} 
{\blue We observe that \eqref{eq:fredholm_eig_problem} has a countably infinite set of solutions consisting of orthogonal POD modes $\varphi_j$ and corresponding POD eigenvalues $\lambda_j$.}
This allows for an expansion of the random variable $q(y)$ as
\begin{align}
    q(y) = \sum_{j=1}^\infty a_j \varphi_j(y) \, ,
\end{align}
where the stochastic coefficients are calculated as $a_j = (q,\varphi_j)_\Omega$.
{\blue This expansion is optimal in terms of capturing the turbulent kinetic energy of the velocity component being considered and the contribution from each mode $\varphi_j$ to the total turbulent kinetic energy contained in this component is given by the POD eigenvalues $\lambda_j$.}
Further, we can also expand the two-point correlation tensor $C(y,y')$ similar to the random variable
\begin{align}
    C(y,y') = \sum_{j=1}^\infty \lambda_j \varphi_j(y) {\blue \varphi_j(y')} \, .
\end{align}
Note that for the subsequent analysis in \S \ref{sec:results}, we consider the two-point correlation of the streamwise velocity fluctuations which we will denote as $C_u$.
The remaining wall-normal and spanwise velocity components are considered in Appendix \ref{App:uni_spanwise_wall_normal} and their two-point correlations will be denoted as $C_v$ and $C_w$, respectively.  

In practice, when only a discrete version of the two-point correlation is known, the POD analysis is performed by a numerical solution of the Fredholm eigenvalue problem in \eqref{eq:fredholm_eig_problem}.
Details of the numerical solution are not discussed here for brevity, but they are included in Appendix \ref{App:Num_solv_POD_eig_prob} for completeness.

\subsection{Degrees of {\blue similarity} in POD}
\label{subsec:similarity_in_POD}

We distinguish between two degrees of Reynolds number {\blue similarity} in the POD analysis.
The first, which we will refer to as strong {\blue similarity}, requires that the Reynolds number dependence of the two-point correlation factors out completely
\begin{equation}
    C(y,y',Re_\tau) = g(Re_\tau) \widetilde{C}(y,y')  \, .
    \label{eq:strong_similarity}
\end{equation}
In this case, the Reynolds number dependence of the two-point correlation is contained entirely within the multiplicative factor $g(Re_\tau)$. Therefore, all the POD modes will be {\blue show similarity} with respect to the Reynolds number and all of the POD eigenvalues will have the same Reynolds number dependence.
This will imply that all the modes $\varphi_j$ in \eqref{eq:PODy} {\blue show similarity}.
However, we note that even if the strong {\blue similarity} in  \eqref{eq:strong_similarity} is not fully satisfied, some degree of {\blue similarity} may still be expected.
Specifically, the second degree of {\blue similarity} which we consider is that of the leading POD modes
\begin{equation}
    \varphi_j(y,Re_\tau) = \widetilde{\varphi}_j(y) \, , \qquad j=1,2,\ldots,N \, ,
    \label{eq:weak_similarity}
\end{equation}
for some finite number $N$, which will imply that only the first few modes $\varphi_j$ in \eqref{eq:PODy} {\blue show similarity}.
We will refer to this as weak {\blue similarity.}
In this case, we also get POD modes {\blue which show similarity}, but the Reynolds number dependence of the POD eigenvalues is allowed to vary from mode to mode.
We argue that weak {\blue similarity} is more relevant than strong {\blue similarity} as any practical turbulence modeling must necessarily involve quite severe truncations, and thus, it is only the leading POD modes that actually matter.

{\blue We emphasize that universality and similarity of statistics in wall-bounded turbulent flows have been studied for a long time by many different authors.
  This includes universality of the logarithmic law of the wall, similarity of the wake region, similarity of the variance of the streamwise velocity fluctuations, universality of the small scale motions, and so on.\cite{marusic2013logarithmic,meneveau2013generalized,yang2018hierarchical,luchini2017universality,yang2018hierarchical_scalar,yang2019hierarchical,lee2019spectral,Smits_et_al_Ret_scale_inner_layer_2021,Chen_et_al_Ret_scale_inner_layer_2021}
  It should therefore be acknowledged that the results on strong and weak similarities presented in this work do not represent a fundamentally new discovery as they are likely manifestations of the same underlying flow similarities observed in these previous works.
  Still, we note that similarity of one-point statistics does not necessarily guarantee similarity of two-point statistics as well.
  For example, the attached eddy hypothesis has implications on both one- and two-point statistics (among others), but the two-point statistics follow the predictions due to the attached eddy hypothesis less closely than one-point statistics. \cite{yang2019hierarchical,Yang2017Structure}
  Therefore, as previously observed similarities primarily concern one-point statistics, the observed strong and weak similarities presented in this work demonstrate that the underlying flow similarities extend to at least some two-point statistics as well, and thus, are more general than previously observed.
  Further, as highlighted by the proposed extension of the LoW in \S III$\,$C, weak universality, in particular, is more relevant for wall modeling in LES than the well-known similarities of one-point statistics.}

\subsection{DNS datasets}
\label{subsec:DNS_channel_datasets}

We make use of plane channel flow DNS for the subsequent analysis.
The data are at four different friction Reynolds numbers $Re_\tau = 180$, $540$, $1000$, and $5200$.
The data for the two lower Reynolds number cases, $Re_\tau = 180$ and $540$, are generated using {\blue a DNS code which has been well validated.\cite{Kim_et_al_1987_DNS_channel}}
The data for the two higher Reynolds numbers, $Re_\tau = 1000$ and $5200$, are obtained from the Johns Hopkins Turbulence Databases (JHTDB).{\blue \cite{Lee_Moser_2015_DNS_chan_5200,Graham_et_al_2016_JHTDB}}
We use a total of 6 snapshots from each of the channel flow cases for the POD analysis.
These snapshots are sampled evenly in time over a full flow-through period.
Additional information about the data used in the POD analysis is given in Table \ref{tab:POD_data_table}.

\bgroup
\def\arraystretch{1.2}
\begin{table}
\centering
\begin{tabular}{ccccc}
\hline
\textbf{\begin{tabular}[c]{@{}c@{}}Reynolds Number:\\ $ \, $\end{tabular}} & \textbf{\begin{tabular}[c]{@{}c@{}}Domain:\\ $L_x^* , L_y^* , L_z^*$\end{tabular}} & \textbf{\begin{tabular}[c]{@{}c@{}}Grid:\\ $N_x , N_y , N_z$\end{tabular}} & \textbf{\begin{tabular}[c]{@{}c@{}}Resolution:\\ $\Delta x^+ , \Delta y_w^+ , \Delta z^+$\end{tabular}} & \textbf{\begin{tabular}[c]{@{}c@{}}\# Snapshots:\\ $ \, $\end{tabular}} \\
 \hline
$Re_\tau = 180$:                                                    & $4\pi , 2 , 2\pi$                                                                  & $192 , 130 , 180$                                                          & $11.8 , 0.01 , 6.3$                                                                                     & 6                                                                       \\
$Re_\tau = 540$:                                                    & $4\pi , 2 , 2\pi$                                                                  & $576 , 244 , 540$                                                          & $11.8 , 0.01 , 6.3$                                                                                     & 6                                                                       \\
$Re_\tau = 1000$:                                                   & $8\pi , 2 , 3\pi$                                                                  & $2048 , 512 , 1536$                                                        & $12.3 , 0.02 , 6.1$                                                                                     & 6                                                                       \\
$Re_\tau = 5200$:                                                   & $8\pi , 2 , 3\pi$                                                                  & $10240 , 1536 , 7680$                                                      & $12.7 , 0.07 , 6.4$                                                                                      & 6 \\                                 \hline                                     
\end{tabular}
\caption{Details about the DNS simulation data used in the POD analysis. Here the asterisk and plus superscripts indicate normalization by the channel half-width $\delta$ and the viscous length scale $\nu / u_\tau$, respectively. Further, the $w$ subscript refers to the value at the wall.}
\label{tab:POD_data_table}
\end{table}
\egroup

To ensure that this database is sufficient for the purposes of this work, we have investigated the convergence of the two-point correlation for the streamwise velocity fluctuations.
{\blue As POD consists of an eigenvalue decomposition of the two-point correlation as discussed in \S \ref{subsec:General_background} above, convergence of the two-point correlation will also ensure convergence of the POD modes and eigenvalues.}
{\blue For all Reynolds numbers, we have calculated a series of two-point correlations using an increasing number of snapshots (from 1 to 6) which are ordered chronologically in time.
The percentage error between neighboring members of this series, e.g., two-point correlations calculated using 3 and 4 snapshots, is then determined by a normalized standard deviation.
The error was found to be less than 1\% between the two-point correlations calculated using 5 and 6 snapshots.
To further assess the convergence, we have calculated an additional series of two-point correlations using 12 snapshots which were also sampled evenly over a full flow-through period (11 for the $Re_\tau = 5200$ case which is the total number available at the JHTDB).
Here, the error was found to already drop below 1\% when comparing the two-point correlations calculated from 6 and 7 snapshots for all Reynolds numbers.
This indicates convergence in roughly a half flow-through period, and thus, justifies the choice of using data covering a full flow-through period for the POD analysis.
Finally, we also compared the error between two-point correlations calculated using 6 and 12 snapshots (6 and 11 for the $Re_\tau = 5200$ case), both sampled evenly in time over a full flow-through period.
The error for this case was found to be around or less than 2\% for all Reynolds numbers.
This demonstrates that the current database is sufficient for the POD analysis and that the results presented below would only show minor changes upon additional convergence.}

\section{Results}
\label{sec:results}

We study the presence or the absence of strong and weak {\blue similarity} of the streamwise velocity fluctuations in \S \ref{subsec:strong} and \S \ref{subsec:weak}, respectively.
A similar analysis for the spanwise and wall-normal components is included in Appendix \ref{App:uni_spanwise_wall_normal} for completeness.
Again, as long as weak {\blue similarity} exists, one can expect an extension of the LoW.
Upon demonstrating weak {\blue similarity}, an extension of the LoW is proposed and assessed in \S \ref{subsec:LoW_extension}.
Finally, we also provide an interpretation of the LoW extension in \S \ref{subsec:interp_LoW_extension}.

\subsection{Strong {\blue similarity}}
\label{subsec:strong}

\begin{figure}
    \centering \includegraphics[width=0.9\textwidth]{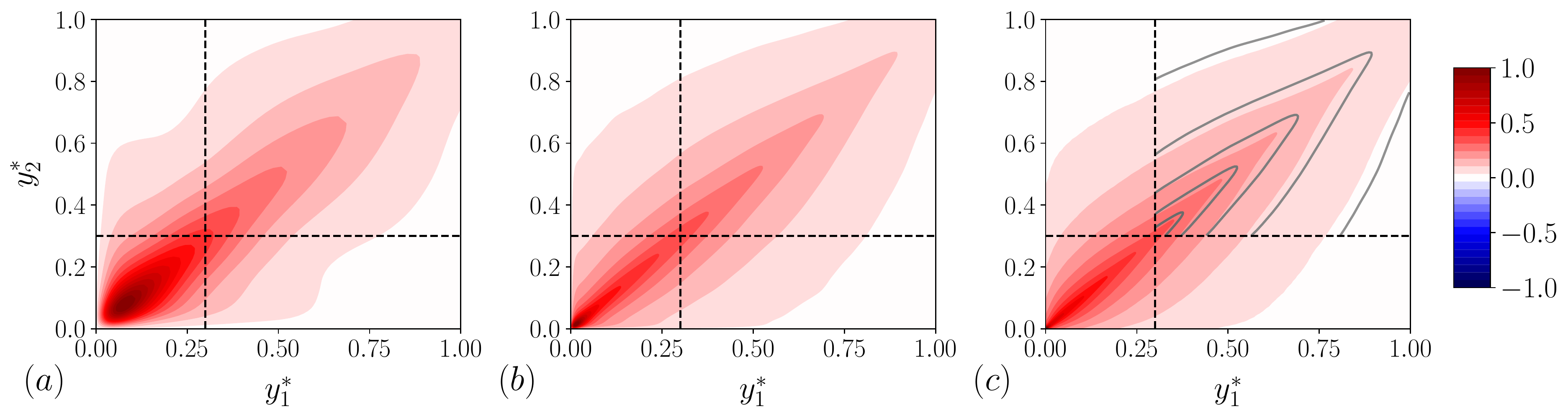}
    \caption{The normalized two-point correlation of the streamwise velocity fluctuations, $C_u$, over the entire half-width of the channel.
    (a), (b), and (c) are for $Re_\tau$ equal to 180, 1000, and 5200, respectively; $y^*=y/\delta$.
    The dashed lines are $y_1^*,y_2^* = 0.3$.
    The gray lines in $(c)$ show the contours from $(b)$ for comparison.}
    \label{fig:Corr_tensor_11_full}
\end{figure}

Using the outer scaling $y^* = y/\delta$,
figure \ref{fig:Corr_tensor_11_full} shows the two-point correlation of the streamwise velocity fluctuation $C_u(y_1^*, y_2^*)$ as a function of $y_1^*$ and $y_2^*$ for channels at $Re_\tau=180$, 1000, and 5200.
The case $Re_\tau = 540$ is excluded here and in the subsequent contour plots for brevity as this two-point correlation is very similar to the $Re_\tau = 1000$ case.
We normalize the data via $\max[C_u(y_1^*,y_2^*)]$ such that the correlation will stay within -1 and 1. 
Note that the plots are symmetric with respect to the line $y_1^* = y_2^*$ due to the commutative law of multiplication.
For any given $y_1^*$ or $y_2^*$, the two-point correlation $C_u$ attains its maximum at $y_1^* = y_2^*$.
A peak emerges in the two-point correlation close to the wall and moves towards the wall as the Reynolds number increases.
This peak corresponds to the peak in $u_{\rm rms}$ and is known to be located at $y_i^+ \approx 15$.{\blue \cite{Pope_Turbulent_Flows_2000}}
Consequently, $y_i^* \approx 15/Re_\tau$ decreases as the Reynolds number increases.
This movement of the peak prevents strong {\blue similarity} over the entire half-width of the channel $0\leq y_1^*,y_2^* \leq 1$. 
However, it does not preclude strong {\blue similarity} in the viscous wall region or wake region separately.

\begin{figure}
    \centering
    \includegraphics[width=0.9\textwidth]{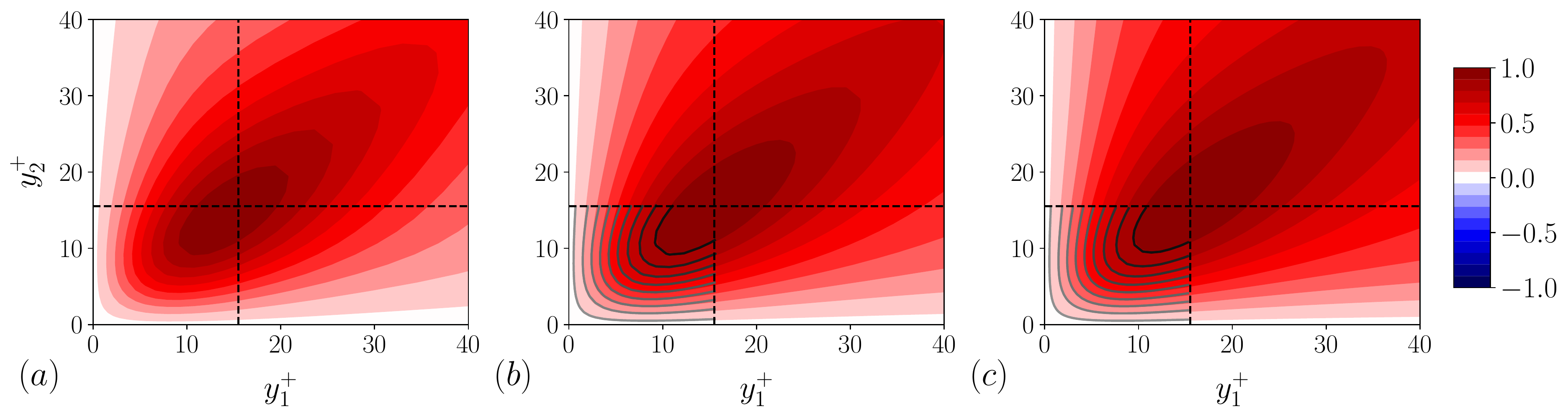}
    \caption{The normalized two-point correlation of the streamwise velocity fluctuations, $C_u$, in the viscous wall region, i.e., for $0\leq y_1^+, y^+_2\leq 40$.
    (a), (b), and (c) are for $Re_\tau$ equal to 180, 1000, and 5200, respectively; $y^+ = (u_\tau/\nu)y$.
    {\blue The dashed lines are $y_1^+,y_2^+ = 15$.
    The gray lines in $(b)$ and $(c)$ show the contours from $(a)$ and (b), respectively, for comparison.}}
    \label{fig:Corr_tensor_11_inner}
\end{figure}

We first focus on the wake region in the interval $0.3 \leq y^* \leq 1$.
By overlaying the two-point correlation at $Re_\tau=1000$ onto that at $Re_\tau=5200$, as done in Figure \ref{fig:Corr_tensor_11_full} (c), we see a good agreement between the two.
This result suggests strong {\blue similarity} in the wake region over the Reynolds number range $Re_\tau = 1000$ to 5200.
Next, we investigate strong {\blue similarity} in the viscous wall region.
Here, we define the viscous wall region to be the layer within which $0 \leq y^+ \leq 40$, where $y^+ = (u_\tau / \nu)y$ is the wall-unit-scaled distance from the wall, for the investigated Reynolds numbers.{\blue \cite{Pope_Turbulent_Flows_2000}}
This definition encompasses both the viscous sublayer and the buffer layer but falls a bit short of the logarithmic layer, i.e., $3\sqrt{Re_\tau}$.{\blue \cite{marusic2013logarithmic}}
An advantage of choosing this specific interval is that it makes the results presented here inter-comparable with and complementary to previous well-known works.{\blue \cite{aubry1988dynamics,Moin_et_al_POD_channel_flow_1989}}
Figure \ref{fig:Corr_tensor_11_inner} shows $C_u$ as a function of $y_1^+$ and $y^+_2$ within $0\leq y_1^+, y_2^+\leq 40$ for flows at $Re_\tau=180$, 1000, and 5200.
We see that there is strong {\blue similarity} in the viscous layer, i.e., over the interval $0 \leq y_1^+, y_2^+ \leq 15$, across the three Reynolds numbers, but no strong {\blue similarity} can be found further away from the wall.
The lack of strong {\blue similarity} outside the viscous layer is expected.
In fact, if we were to plot {\blue $u_{\rm rms}$} without any applied scaling as a function of the wall-normal coordinate, $y^+$, the profiles would not collapse in either the viscous layer or outside the viscous layer.
Here, by re-scaling the two-point correlation per its maximum, there is at least strong {\blue similarity} in the viscous layer.

\subsection{Weak {\blue similarity}}
\label{subsec:weak}

We investigate weak {\blue similarity} in the viscous wall region and the outer part of the logarithmic layer.

\begin{figure}
    \centering   \includegraphics[width=0.9\textwidth]{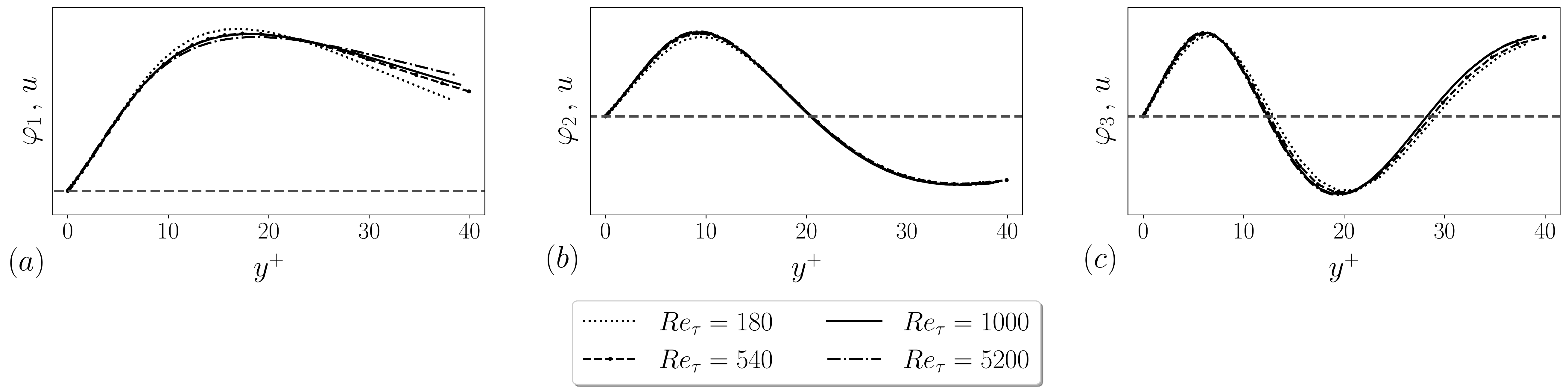}
    \caption{The first three one-dimensional POD modes for the streamwise velocity fluctuations in the viscous wall region, i.e., for $0 \leq y^+ \leq 40$.
    The dashed line is at 0.
    The plots are given with arbitrary units on the ordinate.}
  \label{fig:POD_u_modes_I}
\end{figure}

Figure \ref{fig:POD_u_modes_I} (a-c) shows the first three POD modes of the streamwise velocity fluctuations for $0\leq y^+\leq 40$ at $Re_\tau = 180$, 540, 1000, and 5200.
The four Reynolds numbers give rise to four lines in each plot.
The modes satisfy the no-slip condition at the wall as expected.{\blue \cite{holmes2012turbulence}}
The high-rank modes are more oscillatory than the low-rank modes as observed in other works.{\blue \cite{Moin_et_al_POD_channel_flow_1989,carbone1996hierarchical,baltzer2011structure}}
Importantly, weak {\blue similarity} is observed for the streamwise velocity fluctuations.
That is, the POD modes are Reynolds number independent.
{\blue By definition, the first POD mode contains the most energy.
Therefore, $\varphi_{1, u}$ would be expected to reflect the near-wall peak in $u_{\rm rms}$ at $y^+\approx 15$, as the peak carries the majority of the turbulent kinetic energy in the viscous wall region.
This expectation is confirmed in figure \ref{fig:POD_u_modes_I} (a) where we see that $\varphi_{1, u}$ has a mild peak at $y^+\approx 15$ as well.}

\begin{figure}
    \centering
    \includegraphics[width=0.75\textwidth]{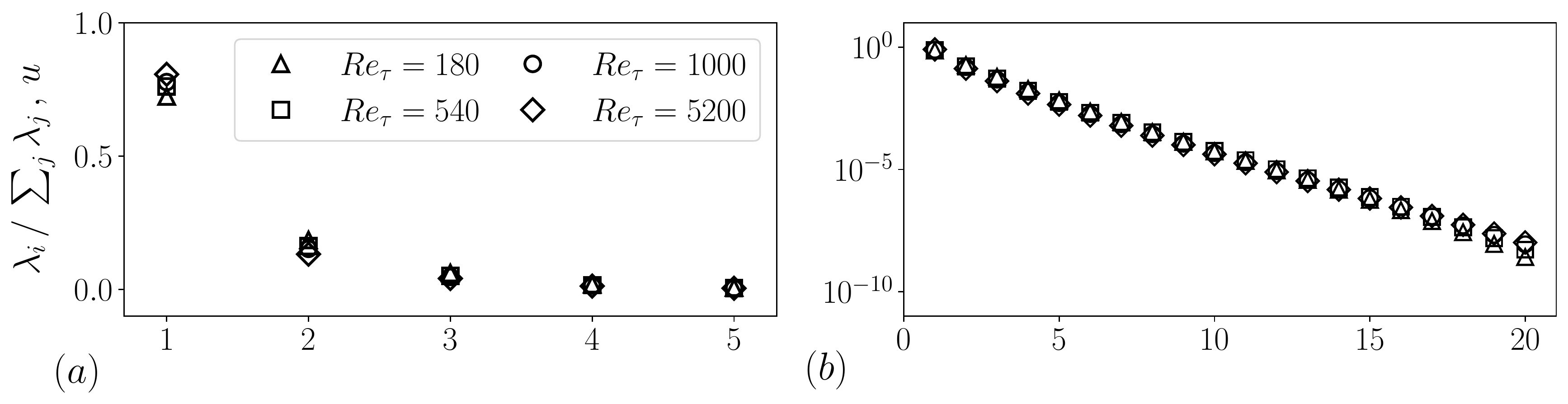}
    \caption{The POD eigenvalues for the streamwise velocity fluctuations within the viscous wall region, i.e., for $0 \leq y^+ \leq 40$.}
    \label{fig:POD_u_eigvals_I}
\end{figure}

Before presenting the results in the logarithmic layer, we comment on our rationale for limiting ourselves to the first three modes.
Figure \ref{fig:POD_u_eigvals_I} shows the POD eigenvalue spectra.
Figure \ref{fig:POD_u_eigvals_I} (a) shows the eigenvalue spectra of the first five modes for the streamwise velocity fluctuations.
We see that the first three POD eigenvalues carry the majority of the total energy at all Reynolds numbers.
In fact, for each of the four Reynolds numbers considered here, the energy in the first three POD modes is $97\%$ or higher.
In addition, the fraction of the energy in each mode remains approximately the same at all Reynolds numbers.
The most noticeable exception to this is the eigenvalues of the first mode.
Per figure \ref{fig:POD_u_eigvals_I} (a), the energy in the first mode increases slightly as the Reynolds number increases.
Figure \ref{fig:POD_u_eigvals_I} (b) shows the eigenvalue spectra of the first twenty modes. 
We observe that the POD eigenvalues decay at an approximately exponential rate.
Further, as no clear deviations from this trend seem to emerge with increasing Reynolds numbers, we expect the picture to remain similar at higher Reynolds numbers.

Next, we investigate weak {\blue similarity} in the outer part of the logarithmic layer.
The logarithmic layer is often defined as spanning somewhere between $y^+ \approx 3\sqrt{Re_\tau}$ and $y^* = O(0.1)$.{\blue \cite{marusic2013logarithmic}}
The definition involves both the inner length scale and the outer length scale, which poses challenges to POD analysis and the subsequent comparative study across different Reynolds numbers.
Here, we will consider the outer part of the logarithmic layer, namely, $0.1 \leq y^* \leq 0.3$, which involves only the outer length scale.
The lower limit $y^*=0.1$ corresponds to $y^+=54$, 100, and 520 at $Re_\tau=$540, 1000, and 5200, which is not too different from the lower limit given by $3\sqrt{Re_\tau}$ at 90, 126, and 288.
Further, we chose to exclude the flow at $Re_\tau=180$ from consideration when making conclusions on weak {\blue similarity} since it is at a low Reynolds number and there is not a convincing logarithmic layer.
Nonetheless, we will still include the POD modes from this case for comparison purposes.

\begin{figure}
    \centering     
    \includegraphics[width=0.9\textwidth]{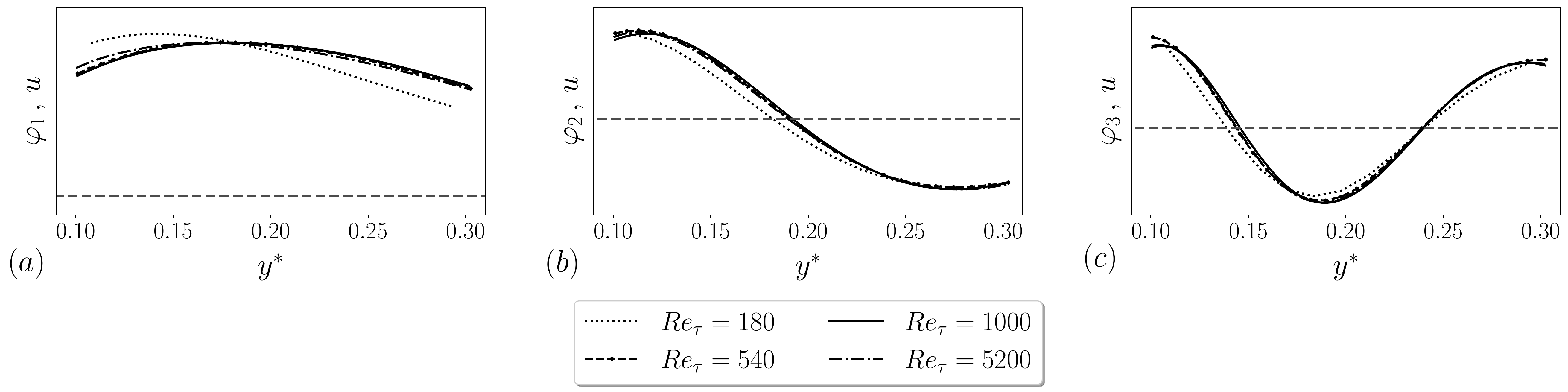}
    \caption{The first three one-dimensional POD modes for the streamwise velocity fluctuations in the outer part of log-layer, i.e., for $0.1 \leq y^* \leq 0.3$.
    The dashed line is at 0.
    The plots are given with arbitrary units on the ordinate.}
    \label{fig:POD_u_modes_L}
\end{figure}

\begin{figure}
    \centering
    \includegraphics[width=0.75\textwidth]{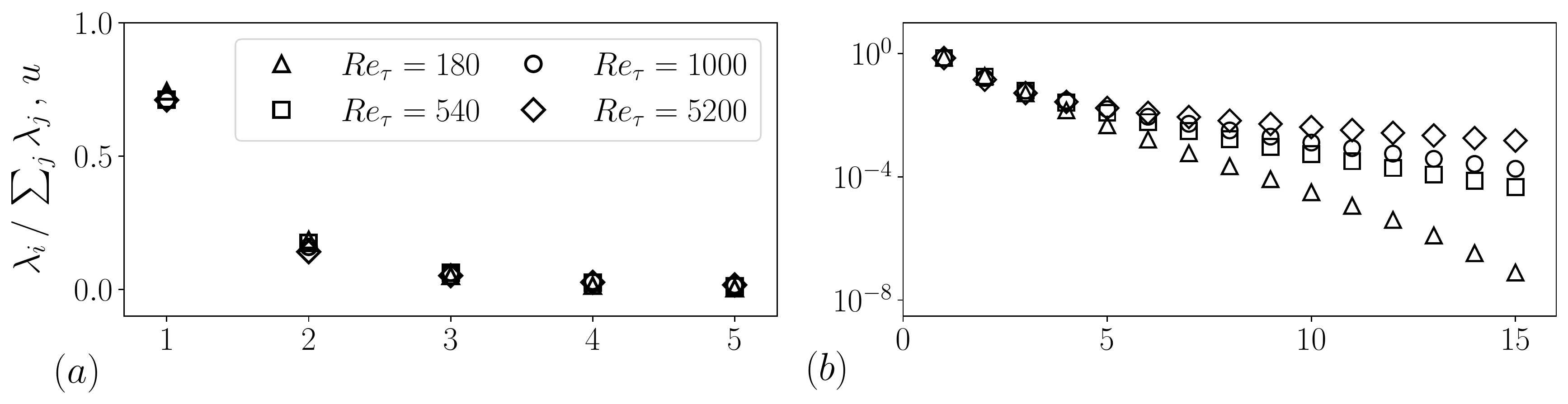}
    \caption{The POD eigenvalues for the streamwise velocity fluctuations in the outer part of log-layer, i.e., for $0.1 \leq y^* \leq 0.3$.}
    \label{fig:POD_u_eigvals_L}
\end{figure}

Figure \ref{fig:POD_u_modes_L} shows the first three POD modes for the streamwise velocity fluctuations at the four Reynolds numbers.
Again, we only show the first three modes because they contain the majority of the total energy.
We observe the following.
Firstly, aside from the $Re_\tau = 180$ modes, there is clear weak {\blue similarity}.
Secondly, the first streamwise mode is approximately constant.
Since the POD modes represent deviations from the mean, this result suggests that the deviation from the LoW in the log-layer region is different from both the LoW itself as assumed in the equilibrium wall model \eqref{eq:LoW} and the linear profile as assumed in \eqref{eq:lineary}. 
{\blue Figure \ref{fig:POD_u_eigvals_L} shows the POD eigenvalue spectra for the streamwise velocity fluctuations in the outer part of the logarithmic layer.
Focusing on figure \ref{fig:POD_u_eigvals_L} (a), we see that the fraction of the total energy contained within the first few modes shows a slight change over the range of Reynolds numbers considered.
Specifically, the energy contained within the first three modes is found to be $97\%$, $95\%$, $93\%$, and $90\%$ for the Reynolds numbers $Re_\tau = 180$, 540, 1000, and 5200, respectively.
This is in contrast to the results from the viscous wall region discussed above where the fraction of energy carried by the first three modes does not decrease with the Reynolds number.
In figure \ref{fig:POD_u_eigvals_L} (b), we see that the POD spectrum shows a large amount of spread between the different Reynolds numbers for the higher eigenvalues. 
Specifically, the higher-order modes carry more energy as the Reynolds number is increased.
Still, even for the highest Reynolds number, the total energy remains dominated by the contributions from the first few modes.}

\subsection{An extension of the law of the wall}
\label{subsec:LoW_extension}

The presence of weak {\blue similarity} indicates the existence of an extension of the LoW.
This suggests that the equilibrium wall model 
\begin{equation}
u({\bf x},t)\approx c(x, z, t){\rm LoW}(y^+) \, ,
\label{eq:EWM}
\end{equation}
where $c$ is a local coefficient, can be extended by the inclusion of an additional term {\blue which shows Reynolds number similarity}.
Here, $\text{LoW}(y^+)$ is the nondimensional law of the wall.
Specifically, ${\rm LoW}(y^+)$ is taken to be the law of the wall function proposed by {\blue Reichardt \cite{reichardt1951vollstandige}} which has been recalibrated to match the mean velocity profile from a $Re_\tau = 5200$ channel.
The analytical expression and recalibrated parameters are given in Appendix \ref{App:LoW_g_formula}.
Thus, we seek a nondimensional term $g$ {\blue which show similarity} such that 
\begin{equation}
    u({\bf x},t) \approx  c_1(x,z,t) \text{LoW}(y^+) + c_2(x,z,t) g(y^+) \, ,
    \label{eq:LoW_extension}
\end{equation}
where $c_1$ and $c_2$ are both local coefficients.

\begin{figure}
    \centering 
    \includegraphics[width=0.75\textwidth]{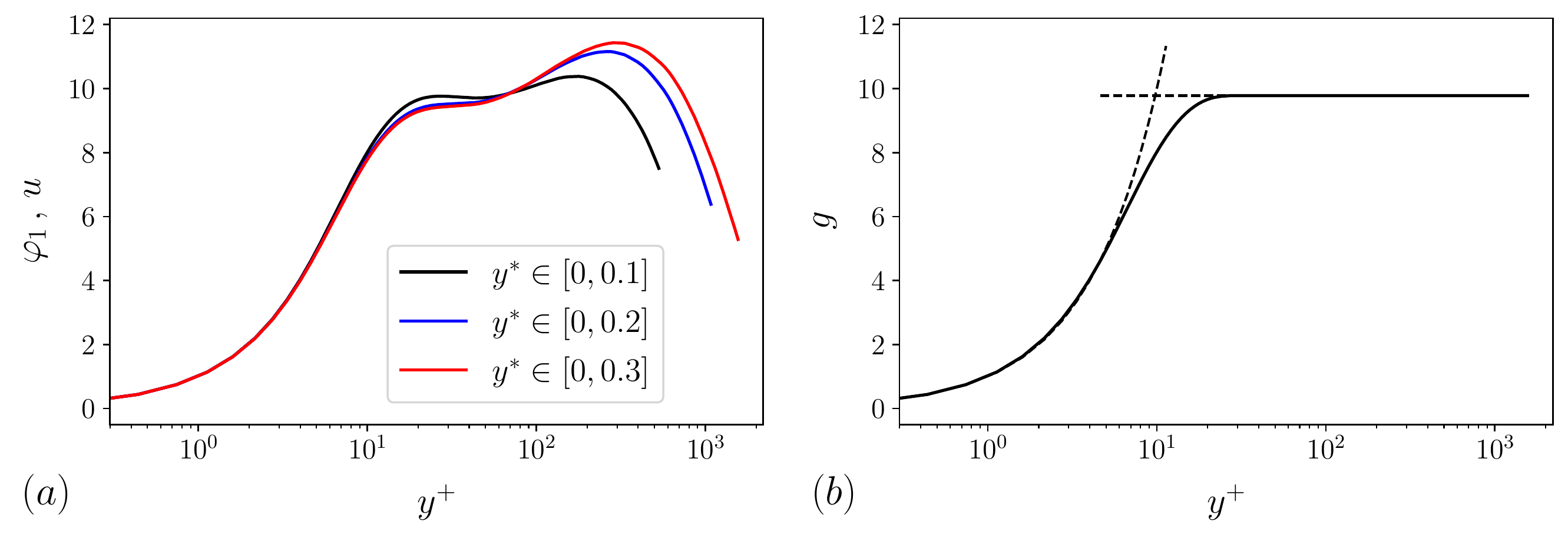}
    \caption{(a) The first streamwise POD mode for $Re_\tau = 5200$ over three wall-normal intervals that start at the wall and end at $y^* = 0.1$, $0.2$, and $0.3$.
    Part (b) shows the $g$ mode. 
    The two dashed lines correspond to $g=y^+$ and $g=9.7$.}
    \label{fig:g_mode_const}
\end{figure}

The exact form of $g$ is not readily available because the results on weak {\blue similarity} in \S\ref{subsec:weak} are confined to either the viscous wall region or the log-layer region, and because they are at finite Reynolds numbers.
To obtain $g$, we therefore repeat the one-dimensional POD analysis over intervals $0 \leq y^* \leq y_\text{max}$ for several $y_\text{max}$.
This exercise will give us insights into possible forms of $g$.
Thanks to weak {\blue similarity}, limiting the analysis to one Reynolds number is warranted.
Furthermore, since near-wall turbulence modeling in LES is intended for flows at high Reynolds numbers, studying a high Reynolds number flow is preferred to a low Reynolds number flow.
Therefore, we perform this new analysis for the $Re_\tau=5200$ case.
The results of this new one-dimensional POD analysis are shown in figure \ref{fig:g_mode_const} (a) in the form of the first POD mode for each interval considered.
We observe that the modes satisfy the no-slip condition. 
They attain a plateau at $y^+\approx 10$, peak further away from the wall, and decrease towards the end of the observation window at $y_{\rm max}$.
The behavior of the POD modes near $y_{\rm max}$ is likely an end effect (similar to the effect of limiting the size of the observation window for a periodic signal).
Considering that the first POD mode in the outer part of the logarithmic layer is approximately constant (see figure \ref{fig:POD_u_modes_L}), a possible mode $g$ is shown in figure \ref{fig:g_mode_const} (b), where the normalization is such that $dg/dy^+=1$ at the wall.
We note that the $g$ mode here is similar to another mode identified in {\blue previous wall modeling efforts.\cite{fowler2022lagrangian}} 
There, the authors were also searching for a non-equilibrium extension of the equilibrium wall model. 
They arrived at such an extension by considering the response of the flow to a suddenly imposed pressure gradient, which features a constant away from the wall and a smooth transition from that constant region to the no-slip wall.
We also provide an analytical expression for the mode $g(y^+)$ in Appendix \ref{App:LoW_g_formula} using a similar functional form as the ${\rm LoW}(y^+)$ function of {\blue Reichardt.\cite{reichardt1951vollstandige}}

We now make use of the extended LoW model in \eqref{eq:LoW_extension} for instantaneous flow reconstructions of the streamwise velocity fluctuations.
Since the mode $g$ in \eqref{eq:LoW_extension} is obtained from POD analysis of the $Re_\tau=5200$ data, the exercise will focus on flow at another Reynolds number, here,  $Re_\tau=1000$.
We note that similar results (not shown) were obtained for both the lower Reynolds number cases $Re_\tau = 180$ and 540, and for the 5200 case.
The reconstructions are for the flow between the wall and $y^*=0.1$, which is the region where wall models are typically applied in LES.
We also include reconstructions from the equilibrium wall model in \eqref{eq:EWM} for comparison purposes.
Creating these reconstructions requires the calculation of coefficients in both \eqref{eq:EWM} and \eqref{eq:LoW_extension}.
The details of these calculations are provided in Appendix \ref{App:Calc_coeffs}.

\begin{figure}
    \centering 
    \includegraphics[width=0.7\textwidth]{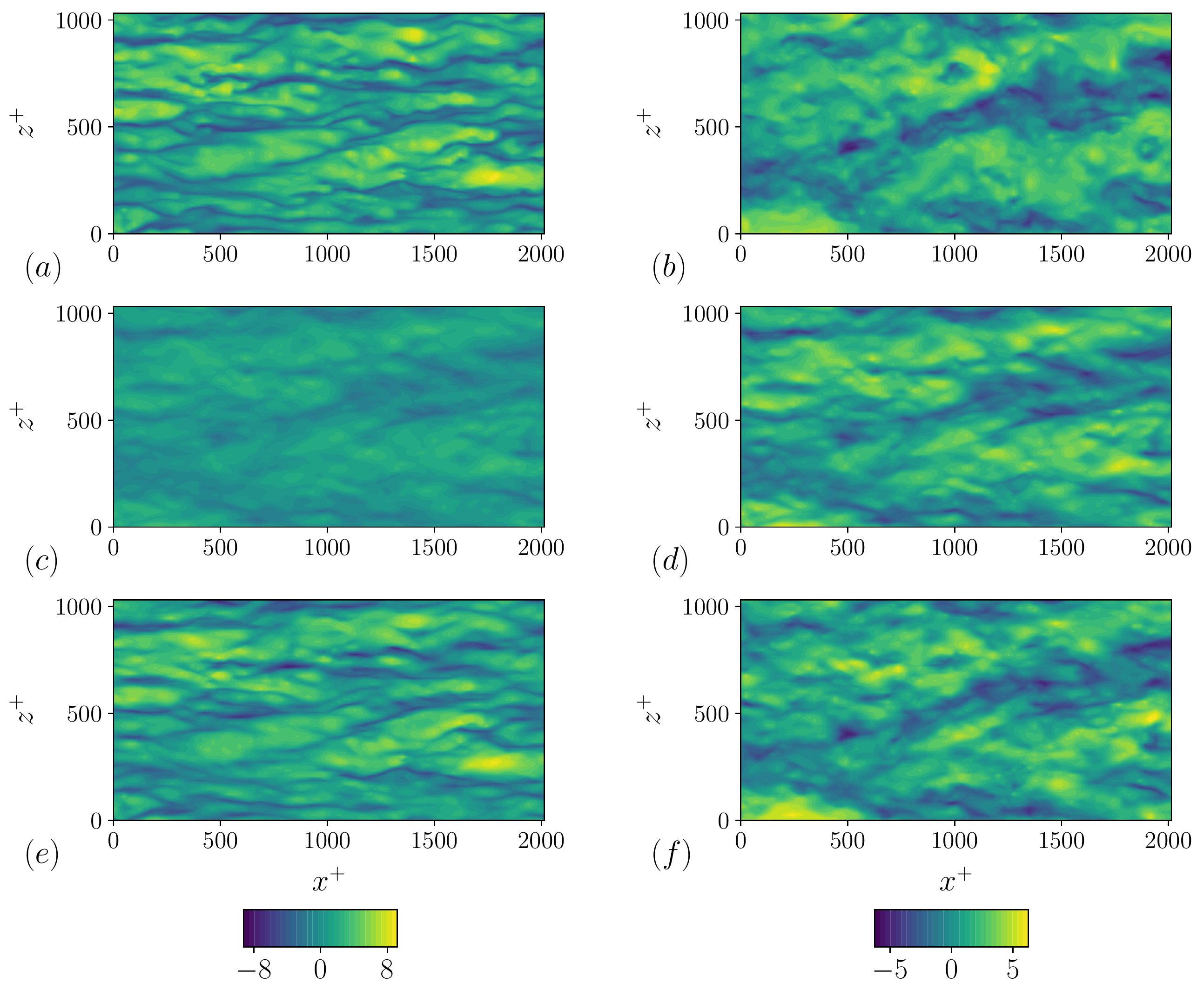}
    \caption{{\blue Streamwise velocity fluctuations $u'$ normalized by the friction velocity $u_\tau$.
    (a, b) DNS data.
    (c, d) Reconstructions according to (15).
    (e, f) Reconstruction according to (16).
    (a, c, e) at $y^+=15$, (b, d, f) at $y^*=0.1$.}}
    \label{fig:u_recon_Ret_1000}
\end{figure}

\begin{figure}
    \centering
    \includegraphics[width=0.8\textwidth]{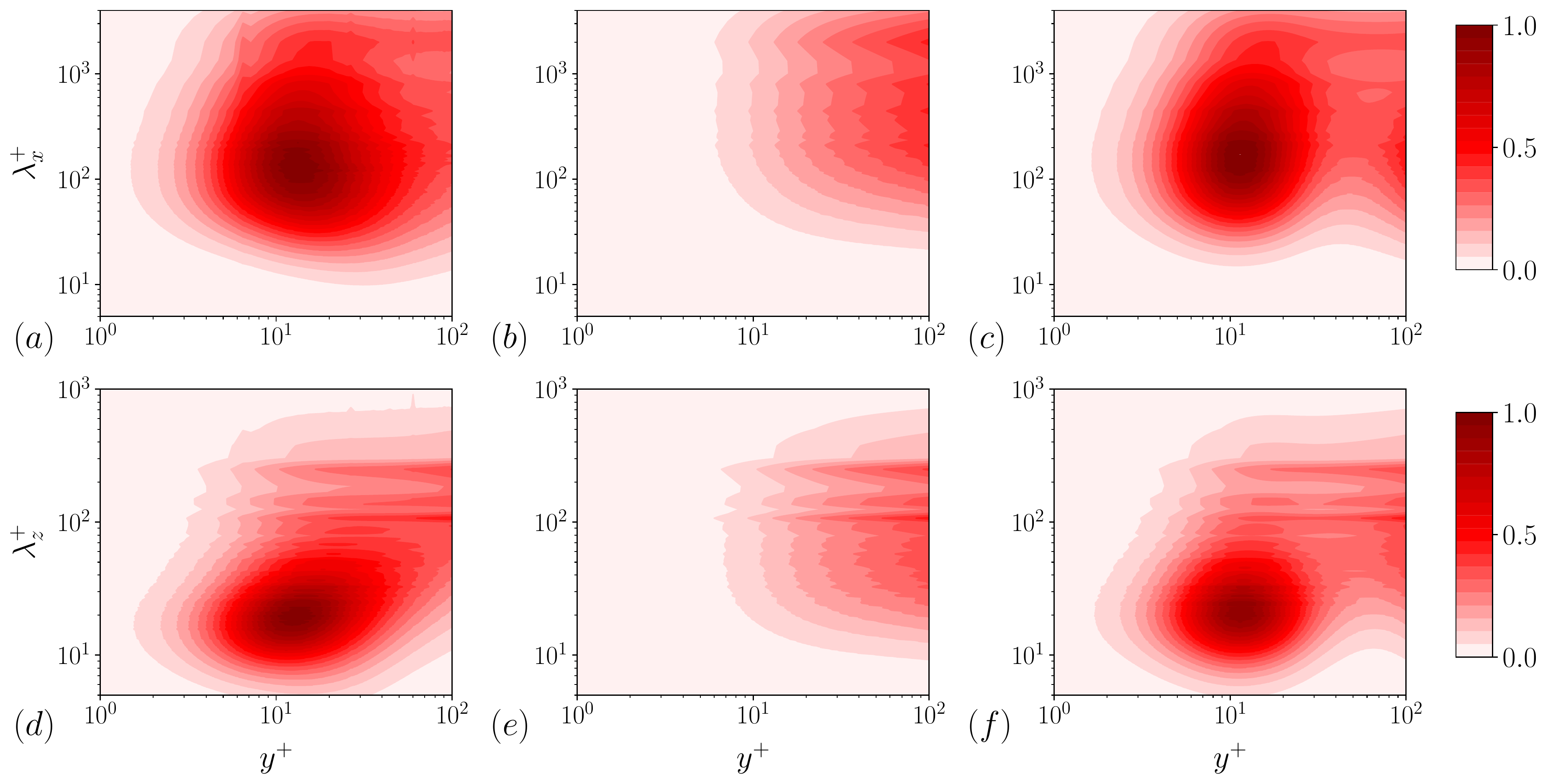}
    \caption{Premultiplied energy spectra of the streamwise velocity fluctuations along the streamwise {\blue $k_x \Phi_{u'u'}^+(y^+,\lambda_x^+)$} and spanwise {\blue $k_z \Phi_{u'u'}^+(y^+,\lambda_z^+)$} directions in wall-unit-scaling.
    (a, b, c) The streamwise spectra from DNS, the equilibrium LoW in \eqref{eq:EWM}, and the extended LoW in \eqref{eq:LoW_extension}, respectively.
    (d, e, f) Same as (a, b, c) but for the spanwise spectra.
    Both the streamwise and spanwise spectra are normalized by the peak value from the DNS spectra.}
    \label{fig:E_u_DNS_and_WM}
\end{figure}

We compare the reconstructions according to \eqref{eq:EWM} and \eqref{eq:LoW_extension} with the DNS data at $Re_\tau = 1000$.
This will serve as a test of the usefulness of the obtained extension of the LoW.
Figure \ref{fig:u_recon_Ret_1000} shows the DNS velocity and the reconstructions according to the equilibrium LoW in \eqref{eq:EWM} and the extended LoW in \eqref{eq:LoW_extension} at $y^+=15$ and $y^*=0.1$.
Comparing figures \ref{fig:u_recon_Ret_1000} (b, d, f), we see that both the equilibrium LoW and the extended LoW give reasonably good reconstructions of the flow field in the log-layer.
Specifically, we observe that the correlation coefficient between the DNS data and the reconstructions from the equilibrium LoW and the extended LoW (calculated using a full snapshot) are 0.75 and 0.84, respectively.
However, comparing figures \ref{fig:u_recon_Ret_1000} (a, c, e), we see that the reconstruction according to the extended LoW provides a much closer agreement with the DNS data than that according to the equilibrium LoW in the near-wall region.
In fact, it is surprising that a reconstruction based on two modes provides almost identical results as the DNS.
Specifically, the reconstruction according to the equilibrium LoW in \eqref{eq:EWM} misses a lot of the small-scale structures that are captured by the extended LoW in \eqref{eq:LoW_extension}.
We note that the correlation coefficient between the DNS data and the reconstructions from \eqref{eq:EWM} and \eqref{eq:LoW_extension} are 0.62 and 0.91, respectively (again calculated using a full snapshot).

Figure \ref{fig:E_u_DNS_and_WM} shows the premultiplied energy spectra of the streamwise velocity fluctuations along the streamwise {\blue $k_x \Phi_{u'u'}^+(y^+,\lambda_x^+)$} and spanwise {\blue $k_z \Phi_{u'u'}^+(y^+,\lambda_z^+)$} directions in wall-unit-scaling.
Figures \ref{fig:E_u_DNS_and_WM} (a, b, c) show the streamwise spectra from DNS, the equilibrium LoW in \eqref{eq:EWM}, and the extended LoW in \eqref{eq:LoW_extension}, respectively, while figures \ref{fig:E_u_DNS_and_WM} (d, e, f) show the same but for the spanwise spectra.
Both the streamwise and spanwise spectra are calculated using data from 96 snapshots which are evenly spaced over a full flow-through period and both have been normalized by the peak value from the DNS spectra.
From the streamwise and spanwise spectra in figure \ref{fig:E_u_DNS_and_WM}, we observe that the near-wall peaks are entirely missed by the equilibrium LoW in \eqref{eq:EWM}, as expected, while the extended LoW in \eqref{eq:LoW_extension} capture the peaks quite accurately, especially given that \eqref{eq:LoW_extension} involves only two modes.
These results further confirm our previous observations in figure \ref{fig:u_recon_Ret_1000};
the reconstruction according to \eqref{eq:LoW_extension} provides a significantly closer agreement with the DNS than that of in \eqref{eq:EWM}.

\subsection{An interpretation of the proposed extension}
\label{subsec:interp_LoW_extension}

\begin{figure}
    \centering 
    \includegraphics[width=0.85\textwidth]{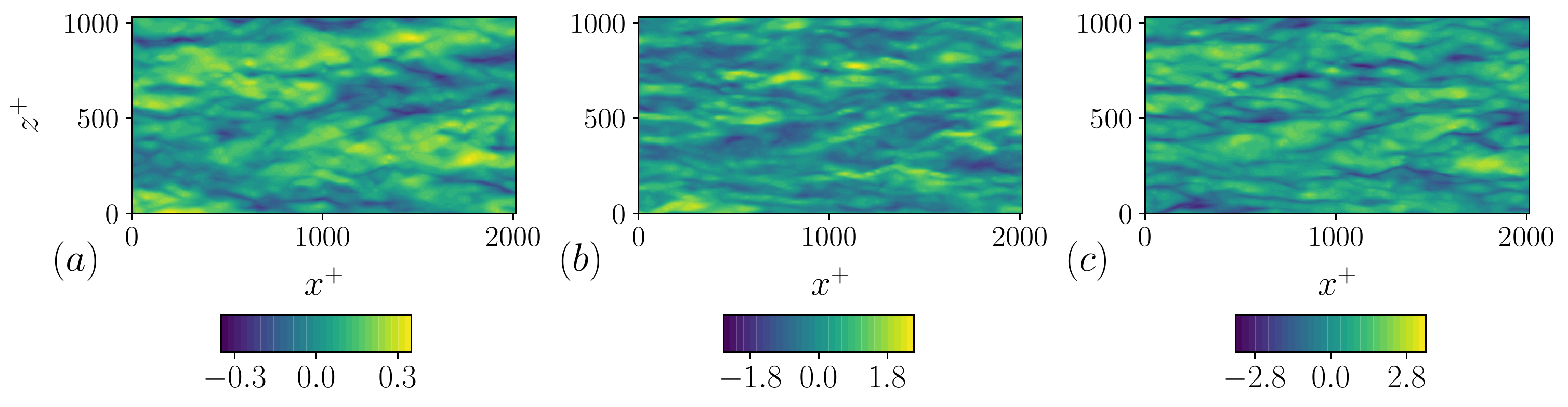}
    \caption{(a) Contours of {\blue $c'(x^+,z^+) / u_\tau$} from \eqref{eq:EWM}.
    (b, c) contours of {\blue $c'_1(x^+,z^+) / u_\tau$} and {\blue $c_2(x^+,z^+) / u_\tau$} from \eqref{eq:LoW_extension}, respectively.
    The flow is at the Reynolds number $Re_\tau = 1000$.}
    \label{fig:EWM_PODWM_Coeffs_Ret_1000}
\end{figure}

\begin{figure}
    \centering 
    \includegraphics[width=0.55\textwidth]{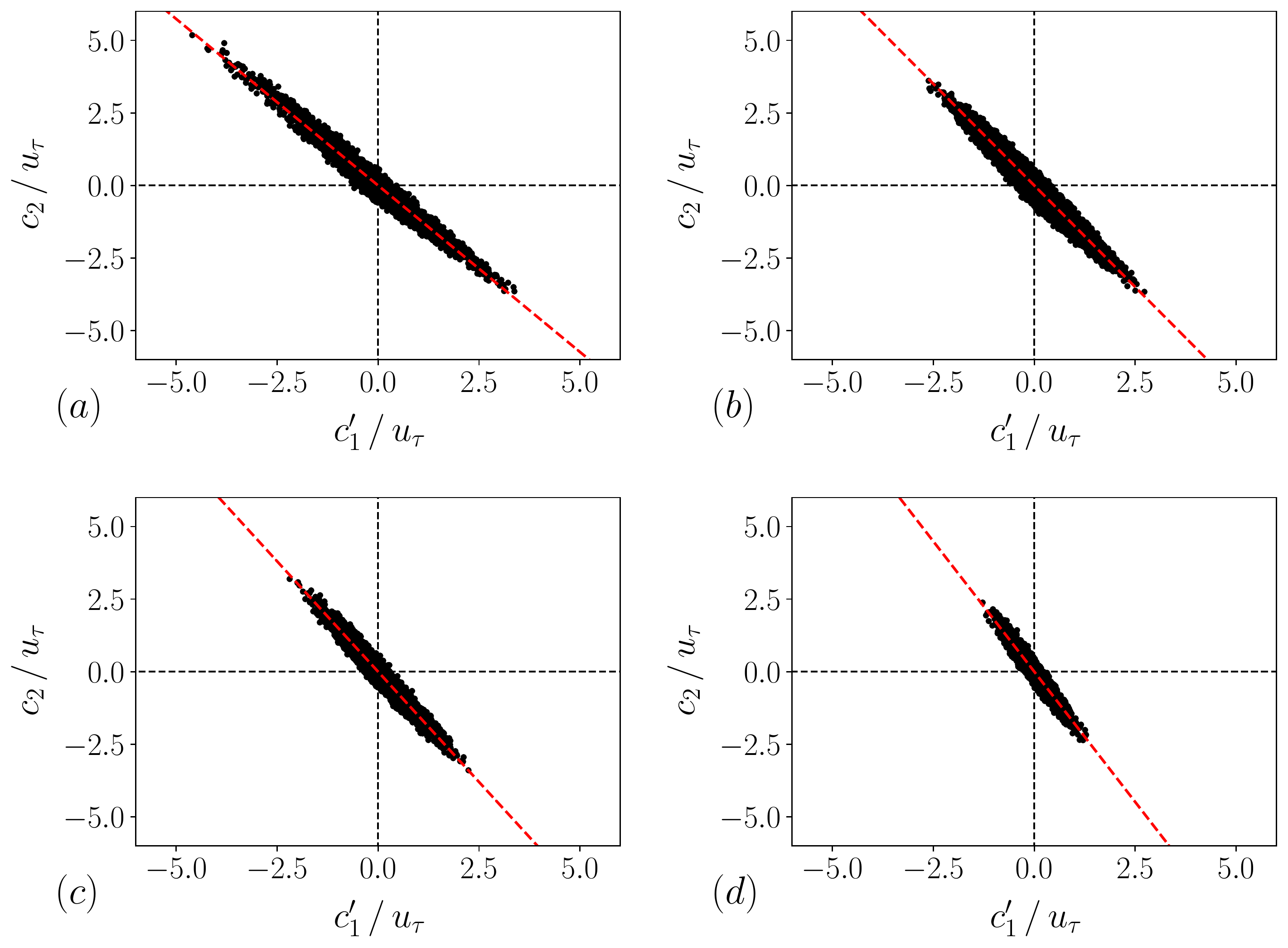}
    \caption{{\blue Scatter plots of $c'_1 / u_\tau$ and $c_2 / u_\tau$} from \eqref{eq:LoW_extension} at (a) $Re_\tau = 180$, (b) 540, (c) 1000, and (d) 5200.
    The red dashed lines show the fit according to \eqref{eq:c1c2}.}
    \label{fig:PODWM_Coeffs_corr_plots}
\end{figure}

\begin{figure}
    \centering 
    \includegraphics[width=0.75\textwidth]{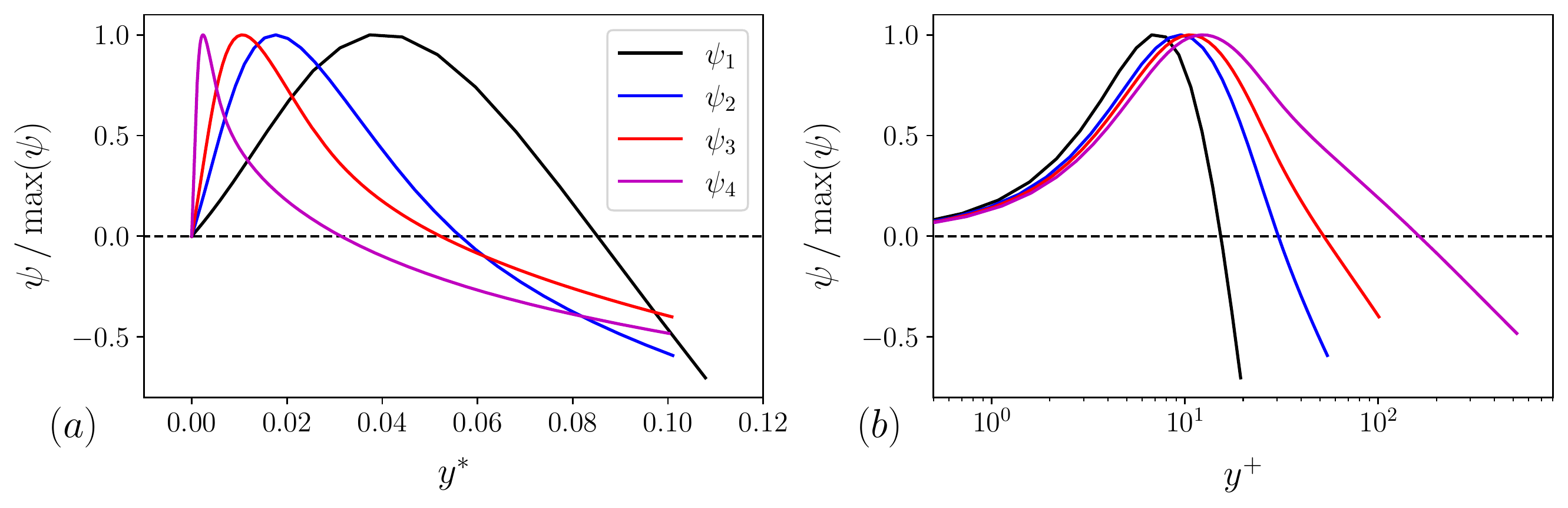}
    \caption{(a) The mode $\psi$ scaled by $\max(\psi)$ as a function of $y^*$ for the Reynolds numbers $Re_\tau = 180$, 540, 1000, and 5200.
    (b) The mode $\psi$ scaled by $\max(\psi)$ as a function of $y^+$ for the same Reynolds numbers.}
    \label{fig:single_mode_plot}
\end{figure}

Having confirmed that the identified extension of the law of the wall in \eqref{eq:LoW_extension} has strong descriptive power, we now move on to provide further analysis and interpretation of the model's behavior.
We first document the behaviors of $c_1$ and $c_2$, i.e., the two coefficients in the extended LoW \eqref{eq:LoW_extension}, and $c$, i.e., the coefficient in the equilibrium LoW \eqref{eq:EWM}, for comparison.
Figure \ref{fig:EWM_PODWM_Coeffs_Ret_1000} shows  $c_1'=c_1-\left<c_1\right>$, $c_2$, and $c'=c-\left<c\right>$ at an arbitrary time instant. 
Here, $\left<c_1\right>=\left<c\right>=u_\tau$ and $\left<c_2\right>=0$ in a channel flow.
We see footprints of streaks in all three plots.
However, the $c'$ contours lack small scales and look more smeared than the $c_1'$ and $c_2$ contours.
Next, figure \ref{fig:PODWM_Coeffs_corr_plots} shows scatter plots of $c'_1 / u_\tau$ and $c_2 / u_\tau$ at the four Reynolds numbers $Re_\tau = 180$, 540, 1000, and 5200.
We observe that, in channel flow, $c'_1$ and $c_2$ are negatively correlated.
{\blue We also note that the rms values of both $c_1'$ and $c_2$ decrease with increasing Reynolds number.
While this might seem somewhat counter-intuitive, it can be explained by the fact that the coefficients are calculated from integrals covering an interval that has a constant length in outer units.
We would, e.g., observe a similar decrease with increasing Reynolds number if the streamwise turbulent kinetic energy was integrated over the interval $0 \leq y^* \leq 0.1$.}
Further, the negative correlation between $c_1'$ and $c_2$ gives the following relation
\begin{equation}
c_2 \approx d(Re_\tau) c_1' \, ,
\label{eq:c1c2}
\end{equation}
where $d(Re_\tau) < 0$ is a Reynolds-number-dependent constant.
Our results show that $d$ equals -1.15, -1.40, -1.52, and -1.80 at $Re_\tau=180$, 540, 1000, and 5200, respectively.
Considering that $\left<u\right> = u_\tau {\rm LoW}(y^+)$ in channel flow, \eqref{eq:LoW_extension} and \eqref{eq:c1c2} implies that
\begin{equation}
u({\bf x},t) \approx u_\tau \text{LoW}(y^+) - a(x,z,t)[{\rm LoW}(y^+) + d(Re_\tau)g(y^+)],
\label{eq:LoW-mixed}
\end{equation}
where $a(x,z,t) = - c_1'(x,z,t)$ and the minus is introduced for convenience in the following.
Comparing \eqref{eq:LoW_extension} and \eqref{eq:LoW-mixed}, we notice that there are two degrees of freedom in \eqref{eq:LoW_extension} whereas there is only one degree of freedom in \eqref{eq:LoW-mixed} (assuming that $d(Re_\tau)$ is known), like in \eqref{eq:EWM}.
This motivates the definition of the following function $\psi(y^+,Re_\tau) = -[{\rm LoW}(y^+) + d(Re_\tau)g(y^+)]$ such that we get $u({\bf x},t) \approx  u_\tau {\rm LoW}(y^+) + a(x,z,t) \psi(y^+,Re_\tau)$.
Figures \ref{fig:single_mode_plot} (a, b) show $\psi$ at the four different Reynolds numbers.
We observe that the $\psi$ mode bears a striking resemblance with the velocity fluctuations due to the wall-attached eddies identified in previous work by other authors\cite{hu2020wall}; it peaks at $y^+\approx 12$ and follows a logarithmic scaling beyond.
This resemblance provides further confidence in the extended LoW in \eqref{eq:LoW_extension}.

\section{Concluding remarks}
\label{sec:conclusion}

This paper discusses the {\blue similarity} of the streamwise velocity fluctuations in turbulent channel flows.
The discussion hinges upon a one-dimensional POD analysis and the concept of strong and weak {\blue similarities}.
Given a wall-normal observational window, strong {\blue similarity} requires {\blue that the two-point correlation show Reynolds number similarity which means that all of the POD modes show similarity as well}.
Here, a strong {\blue similarity} is found for the streamwise velocity fluctuations in the wake region ($0.3 \leq y^* \leq 1$) over the Reynolds number range $Re_\tau = 1000$ to 5200 and in the viscous layer ($0 \leq y^+ \leq 15$) for the Reynolds number range $Re_\tau = 180$ to 5200.
Weak {\blue similarity}, on the other hand, requires that the first or the first few POD modes are {\blue show Reynolds number similarity.}
Such weak {\blue similarity} is found for the streamwise velocity fluctuations within both the viscous wall region ($0 \leq y^+ \leq 40$) and the outer part of the logarithmic layer ($0.1 \leq y^* \leq 0.3$).

The existence of the weak {\blue similarity} suggests the existence of an extension of the LoW (which contains both the viscous layer and the logarithmic layer), and such an extension is proposed based on our one-dimensional POD analysis.
We apply the extended LoW and reconstruct the near-wall flow field over $0 \leq y^* \leq 0.1$.
Compared to the reconstruction according to the equilibrium LoW in \eqref{eq:EWM}, the reconstruction according to the extended LoW in \eqref{eq:LoW_extension} provides a strikingly close agreement with the DNS data.
This is observed from both instantaneous reconstructions of the flow field as well as from premultiplied energy spectra.

Last, we note that this paper is limited to channel flow.
Therefore, future work will extend the current analysis to include pipe and boundary layer flows as well to investigate if the extended LoW would perform equally well for these cases.
Another important point is that channel flow is, on average, at equilibrium.
This motivates further investigation of the application of the extended LoW to non-equilibrium flows to evaluate its performance in this context.
In previous work,{\blue \cite{hansen_yang_abkar_2023}} we have already applied the extended LoW for LES wall modeling, and the model showed superior performance compared with the equilibrium LoW for highly non-equilibrium flows in simple geometries.
Future efforts will therefore be focused on more complicated cases involving complex geometries.
Finally, considering the accurate reconstruction of the wall layer provided by the extended LoW, future investigations will also explore wall-modeled LES of particle-laden flows, for which capturing near-wall intermittency is a critical factor.

\section*{Acknowledgment}

\noindent C.H. and M.A. acknowledge the financial support from the Independent Research Fund Denmark (DFF) under Grant No. 1051-00015B.
X.Y. acknowledges financial support from the US Office of Naval Research contract N000142012315 and Air Force Office of Scientific Research contract FA9550-23-1-0272.
This work was also partially supported by the Danish e-Infrastructure Cooperation (DeiC) National HPC under grant number DeiC-AU-N2-2023005.

\section*{Conflict of interest}

\noindent The authors claim no conflict of interest.

\appendix
\section{{\blue Similarity} of the remaining velocity components}
\label{App:uni_spanwise_wall_normal}
\subsection{Strong {\blue similarity}}

\begin{figure}
    \centering
    \includegraphics[width=0.9\textwidth]{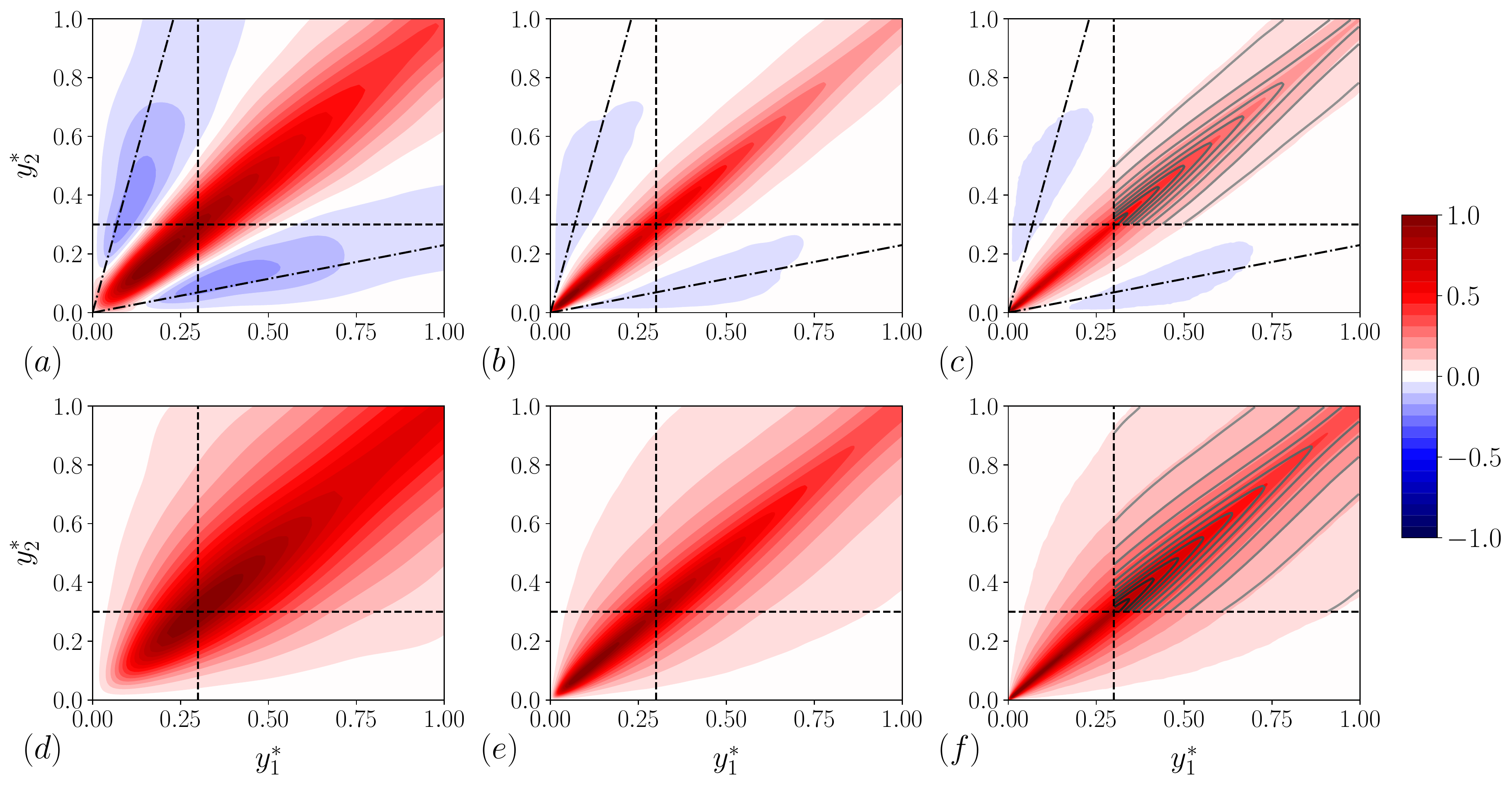}
    \caption{The normalized two-point correlation of the spanwise (a, b, c) and wall-normal (d, e, f) velocity fluctuations, $C_w$ and $C_v$, over the entire half-width of the channel.
    (a, d), (b, e), and (c, f) are for $Re_\tau$ equal to 180, 1000, and 5200, respectively; $y^*=y/\delta$.
    The dashed lines are $y_1^*,y_2^* = 0.3$.
    The gray lines in (c) and (f) show the contours from (b) and (e), respectively, for comparison.}
    \label{fig:Corr_tensor_33_22_full}
\end{figure}

We repeat the exercise in \S\ref{subsec:strong} concerning strong {\blue similarity} for the spanwise and the wall-normal components.
The results over the full half-width of the channel are shown in figure \ref{fig:Corr_tensor_33_22_full}.
The following observations can be made.
Firstly, both $C_w$ and $C_v$ attain their maximum close to the wall, like $C_u$.
However, the wall-normal ranges within which $C_w$ and $C_v$ take large values are much more extended than that in $C_u$, with $C_v$ staying large over an even more extended wall-normal distance range than $C_w$.
Secondly, negative correlations are found in $C_w$ along $y_1 \approx 0.23y_2$ and $y_1 \approx 4.3 y_2$.
Comparing figures \ref{fig:Corr_tensor_33_22_full} (b, c), we see that these areas of negative correlations do not reduce in size with increasing Reynolds number, suggesting that the flow structures responsible for these negative correlations scale with the outer length scale.
Thirdly, as the results at $Re_\tau=1000$ and 5200 collapses in the wake region, there is strong {\blue similarity} for both the spanwise and wall-normal velocity fluctuations in the wake region over the Reynolds number range $Re_\tau = 1000$ to 5200.
This is consistent with outer layer similarities in both $w_{\rm rms}$ and $v_{\rm rms}$.{\blue \cite{Lee_Moser_2015_DNS_chan_5200,yang2019hierarchical}} 

\begin{figure}
    \centering
    \includegraphics[width=0.9\textwidth]{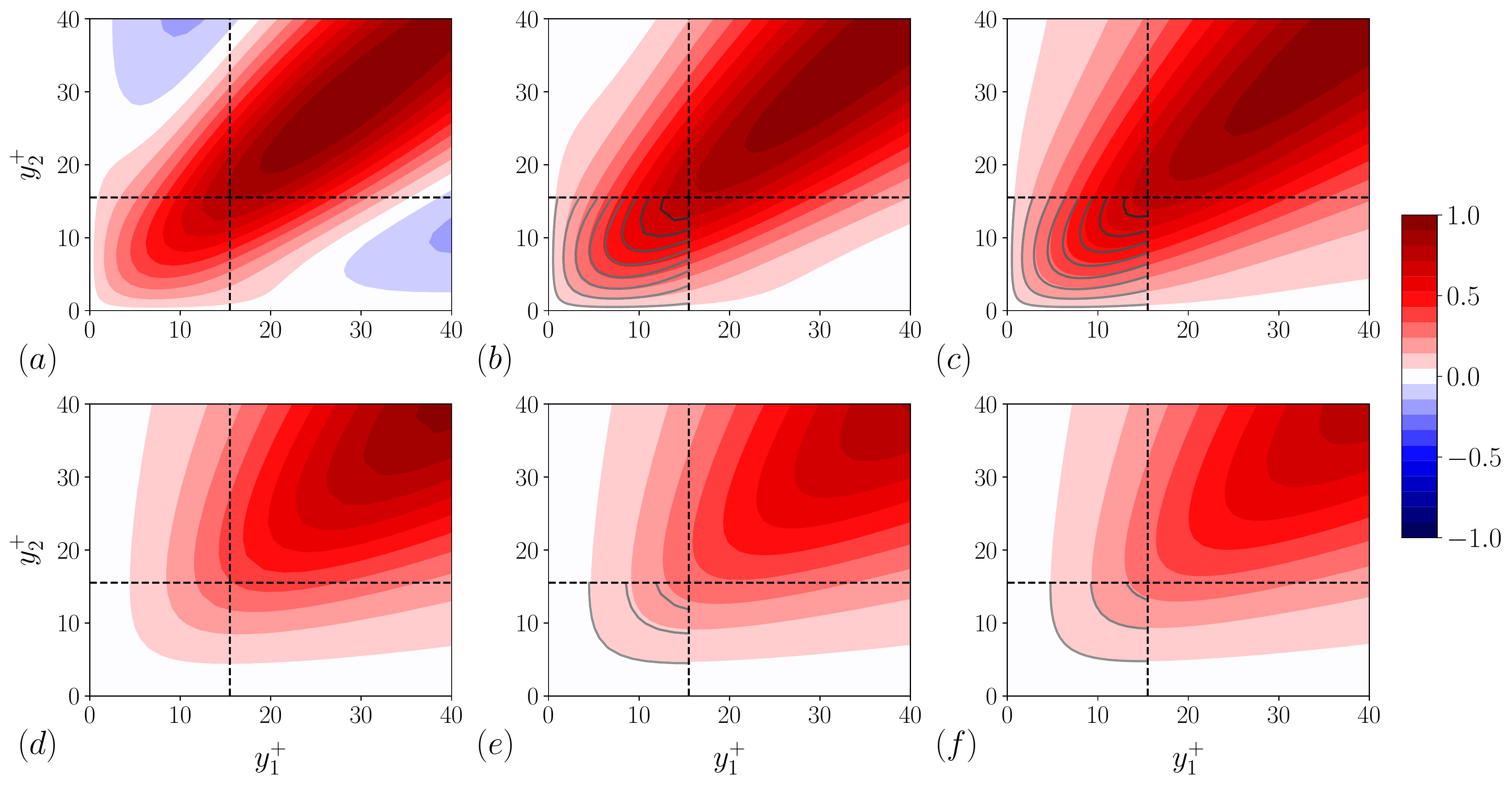}
    \caption{The normalized two-point correlation of the spanwise (a, b, c) and wall-normal (d, e, f) velocity fluctuations, $C_w$ and $C_v$, in the viscous wall region, i.e., for $0\leq y_1^+, y^+_2\leq 40$.
    (a, d), (b, e), and (c, f) are for $Re_\tau$ equal to 180, 1000, and 5200, respectively; $y^+ = (u_\tau/\nu)y$.
    {\blue The dashed lines are $y_1^+,y_2^+ = 15$.
    The gray lines in (b) and (c) show the contours from (a) and (b), respectively, while the gray lines in (e) and (f) show the contours from (d) and (e), respectively, for comparison.}}
    \label{fig:Corr_tensor_33_22_inner}
\end{figure}

Repeating the exercise above, we show the spanwise and the wall-normal two-point correlations $C_w$ and $C_v$ for $0\leq y_1^+, y^+_2\leq 40$ in figure \ref{fig:Corr_tensor_33_22_inner}.
We observe the following.
Firstly, from figure \ref{fig:Corr_tensor_33_22_inner} (b, c), we see that strong {\blue similarity} for the spanwise component is not observed over the full viscous wall region ($0 \leq y^+ \leq 40$), but that it is observed in the viscous layer ($0 \leq y^+ \leq 15$).
This is similar to the streamwise component.
Secondly, {\blue the regions of negative correlation in figure \ref{fig:Corr_tensor_33_22_inner} (a) seem to match well with a previously identified dominant coherent motion in a turbulent channel.\cite{Moin_et_al_POD_channel_flow_1989}}
This coherent motion consists of a streamwise vortex pair centered at around $y^+ = 30$, which gives rise to an anti-correlated spanwise velocity below and above $y^+=30$.
Thirdly, strong {\blue similarity} is present in the viscous layer for the wall-normal component, but not over the full viscous wall region, as can be seen from figure \ref{fig:Corr_tensor_33_22_inner} (e, f).

\subsection{Weak {\blue similarity}}

We investigate weak {\blue similarity} in the near-wall region and the outer part of the log-layer.

\begin{figure}
    \centering
    \includegraphics[width=0.9\textwidth]{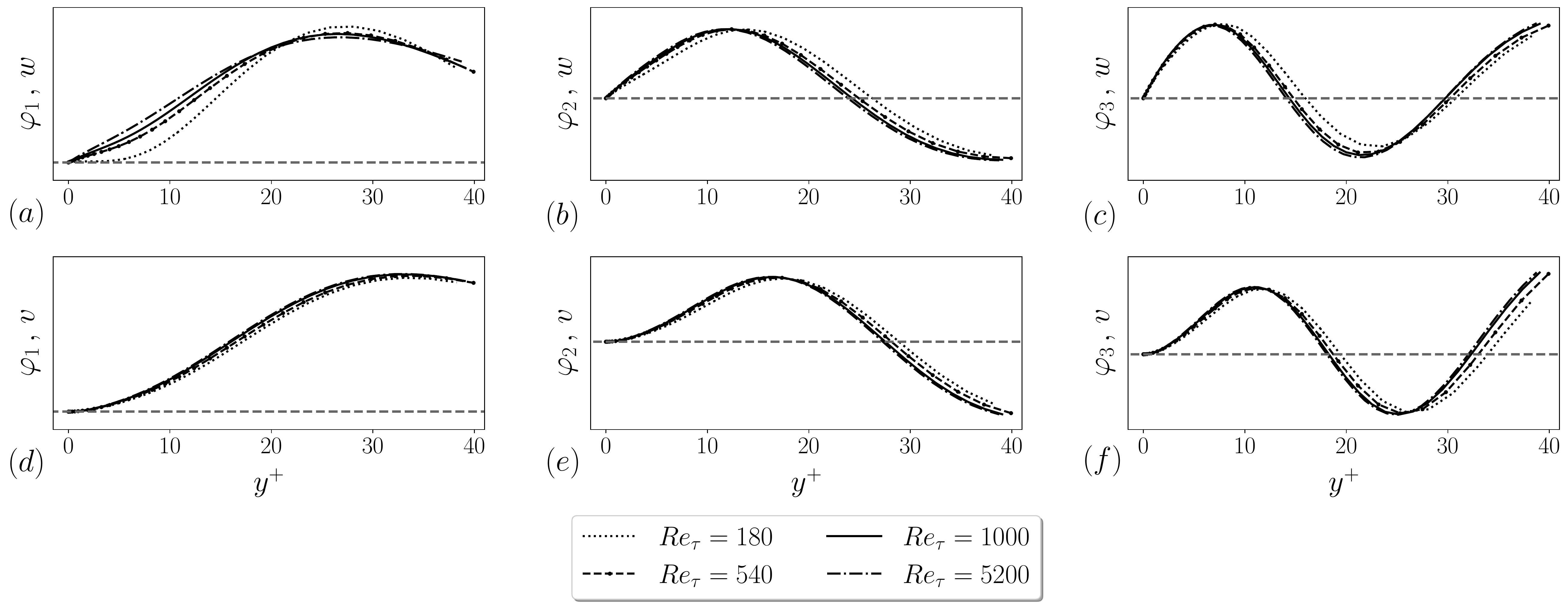}
    \caption{The first three one-dimensional POD modes for the spanwise (a, b, c) and wall-normal (d, e, f) velocity fluctuations in the viscous wall region, i.e., for $0 \leq y^+ \leq 40$.
    The dashed line is at 0.
    The plots are given with arbitrary units on the ordinate.}
  \label{fig:POD_w_v_modes_I}
\end{figure}

Figures \ref{fig:POD_w_v_modes_I} (a-f) show the first three POD modes (columns) of the spanwise and wall-normal velocity fluctuations (rows) for $0\leq y^+\leq 40$ at $Re_\tau=180$, 540, 1000, and 5200.
We observe weak {\blue similarity} for both the spanwise and wall-normal cases.
The only noticeable deviation from this weak {\blue similarity} is the spanwise POD modes at $Re_\tau = 180$, which is probably a low Reynolds number effect.
It is also intriguing that the POD modes of different velocities have similar characteristics (see figure \ref{fig:POD_u_modes_I} for the streamwise case).
Similar to the streamwise case in figure \ref{fig:POD_u_modes_I}, the first spanwise and wall-normal POD modes show a signature of the peak in the corresponding rms profiles.
Specifically, the spanwise and wall-normal velocity rms' attain their maximum further away from the wall than for the streamwise case, and the mild peaks in $\varphi_{1, w}$ and $\varphi_{1, v}$ are correspondingly further away from the wall.

\begin{figure}
    \centering
    \includegraphics[width=0.9\textwidth]{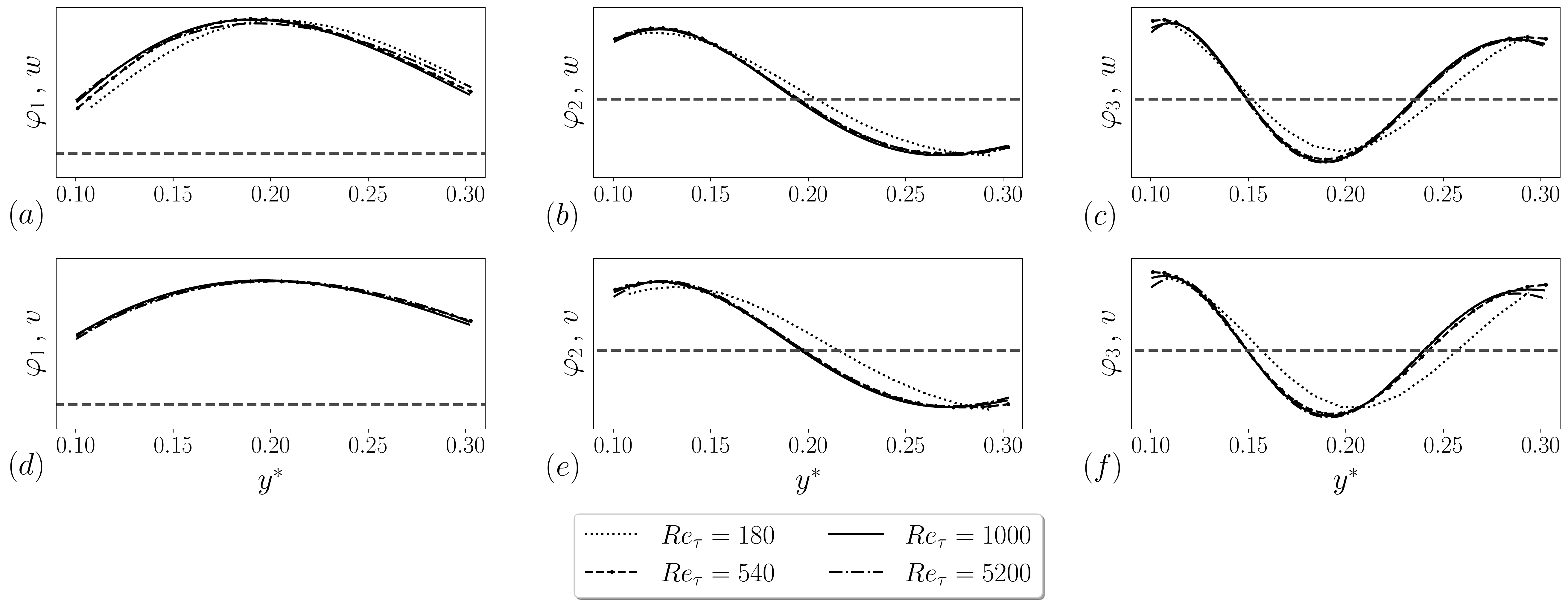}
    \caption{Same as figure \ref{fig:POD_w_v_modes_I} but for $0.1\leq y^*\leq 0.3$.}
  \label{fig:POD_w_v_modes_L}
\end{figure}

Next, we investigate weak {\blue similarity} in the outer part of the log-layer which we take as $0.1 \leq y^* \leq 0.3$ as discussed in \S \ref{subsec:weak}.
Figure \ref{fig:POD_w_v_modes_L} shows the first three POD modes for the spanwise and wall-normal velocity fluctuations at the four Reynolds numbers.
Firstly, there is clear weak {\blue similarity} for both the spanwise and wall-normal cases, aside from the $Re_\tau = 180$ modes.
Secondly, the modes are again similar between the different velocity components (see figure \ref{fig:POD_u_modes_I} for the streamwise case).

\section{Numerical solution of the POD eigenvalue problem}
\label{App:Num_solv_POD_eig_prob}

We consider the one-dimensional scalar variant of the POD eigenvalue problem which is restated here for convenience
\begin{equation}
    \int_\Omega C(y,y') \varphi(y') \, dy' = \lambda \varphi(y) \, .
\end{equation}
A discretization of this problem onto the grid points $y_j$ for $j=1,\ldots,N$ can be performed as follows.
First, we approximate the integral using a quadrature rule
\begin{equation}
    \int_\Omega f(y) \, dy = \sum_{k=1}^N \omega_k f(y_k) \, ,
\end{equation}
where $\omega_k$ are the weights for the particular quadrature method considered.
Following {\blue previous work from other authors,\cite{Moin_et_al_POD_channel_flow_1989}} we use the trapezoid rule for this discretization which is justified because two-point correlations, being statistical objects, are naturally quite smooth.
Thus, for each grid point $y_j$, we get
\begin{equation}
    \sum_{k=1}^N \omega_k C(y_j,y_k) \varphi(y_k) = \lambda \varphi(y_j) \, , \qquad j=1,\ldots,N \, .
\end{equation}
Putting this into matrix form results in a discrete eigenvalue problem
\begin{equation}
    \boldsymbol C \boldsymbol D \boldsymbol \varphi = \lambda \boldsymbol \varphi \, ,
\end{equation}
where we have introduced the two matrices
\begin{equation}
    \boldsymbol C =
    \left[
    \begin{array}{ccc}
        C(y_1,y_1) & \cdots & C(y_1,y_N) \\
        \vdots & \ddots & \vdots \\
        C(y_N,y_1) & \cdots & C(y_N,y_N)
    \end{array}
    \right]
    \, , \qquad 
    \boldsymbol D = 
    \left[
    \begin{array}{ccc}
        \omega_1 & & \\
        & \ddots & \\
        & & \omega_N
    \end{array}
    \right]
    \, .
\end{equation}
For uniform grids, the matrix $\boldsymbol C \boldsymbol D$ will be symmetric and the POD eigenvalue problem can be solved directly.
However, for non-uniform grids, additional steps are required.{\blue \cite{Moin_et_al_POD_channel_flow_1989,smith2005low}}
In this case, the matrix $\boldsymbol C \boldsymbol D$ is no longer symmetric (meaning that real eigenvalues are not guaranteed), and thus, a rescaling of the problem must be performed.
This rescaling can be done by multiplying the eigenvalue problem with $\boldsymbol D^{1/2}$ from the left, while we also move a factor $\boldsymbol D^{1/2}$ from $\boldsymbol C \boldsymbol D$ over to the mode $\boldsymbol \varphi$.
This gives the rescaled eigenvalue problem
\begin{equation}
    \boldsymbol C_D \boldsymbol \varphi_D = \lambda \boldsymbol \varphi_D \, ,
\end{equation}
where we have introduced the notation
\begin{equation}
    \boldsymbol C_D = \boldsymbol D^{1/2}  \boldsymbol C  \boldsymbol D^{1/2} \, , \qquad \boldsymbol \varphi_D = \boldsymbol D^{1/2} \boldsymbol \varphi \, .
\end{equation}
Thus, one solves this eigenvalue problem to obtain the eigenvalues $\lambda$ and modes $\boldsymbol \varphi_D$, after which the value of the POD modes at the grid points can be computed as 
\begin{equation}
    \boldsymbol \varphi = \boldsymbol D^{-1/2} \boldsymbol \varphi_D \, .
\end{equation}

\section{Analytical expressions for the extended LoW modes}
\label{App:LoW_g_formula}

\begin{figure}
    \centering 
    \includegraphics[width=0.75\textwidth]{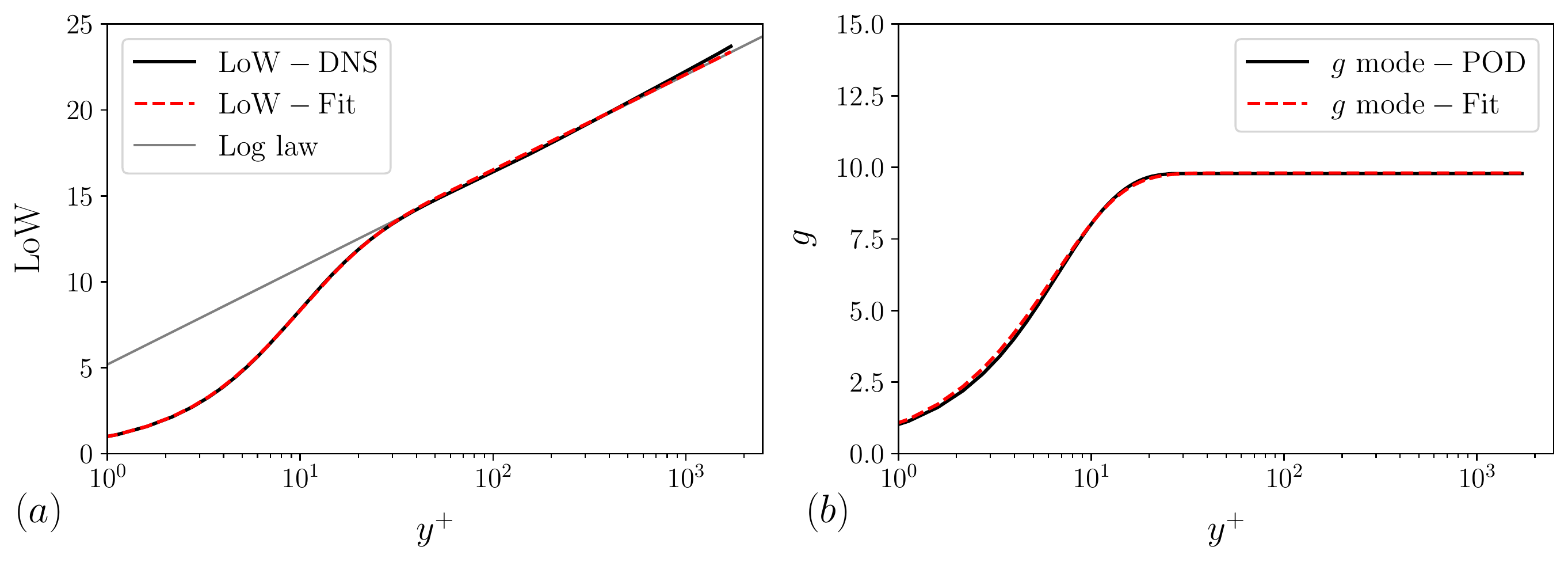}
    \caption{(a) Comparison of the ${\rm LoW}$ mode from DNS at $Re_\tau = 5200$ and from \eqref{eq:LoW_formula}.
    (b) Comparison of the $g$ mode from POD at $Re_\tau = 5200$ and from \eqref{eq:g_formula}.}
    \label{fig:LoW_and_g_modes}
\end{figure}

Here, we provide analytical expressions for the ${\rm LoW}(y^+)$ and $g(y^+)$ modes in \eqref{eq:EWM} and \eqref{eq:LoW_extension}.
For the ${\rm LoW}$ mode, there are several options.{\blue \cite{reichardt1951vollstandige,spalding1961single}}
We choose to consider the following expression from {\blue Reichardt \cite{reichardt1951vollstandige}} given by 
\begin{align}
    {\rm LoW}(y^+) = \frac{1}{\kappa}\ln(1 + \kappa y^+) + A_1\left[ 1 - \exp\left(-\frac{y^+}{A_2}\right) - \left( \frac{y^+}{A_2}\right)\exp\left(-A_3y^+\right)  \right] \, ,
    \label{eq:LoW_formula}
\end{align}
where $\kappa = 0.41$ is the von Kármán constant.
We have recalibrated the constants $A_1$, $A_2$, and $A_3$ by fitting the above expression to the mean velocity profile {\blue over the interval $0 \leq y^+ \leq 1000$} in a $Re_\tau = 5200$ channel.{\blue \cite{Lee_Moser_2015_DNS_chan_5200}}
The resulting values are $A_1 = 7.4$, $A_2 = 9.5$, and $A_3 = 0.29$.
Figure \ref{fig:LoW_and_g_modes} (a) shows a comparison of the ${\rm LoW}$ mode from DNS at $Re_\tau = 5200$ with that from \eqref{eq:LoW_formula}.
For the $g$ mode, we have found an analytical expression by fitting a functional form that is similar to the second part of \eqref{eq:LoW_formula} to the mode in figure \ref{fig:g_mode_const}, also over the interval $0 \leq y^+ \leq 1000$.
The resulting expression is
\begin{align}
    g(y^+) = B_1 \left[ 1 - \exp\left(-\frac{y^+}{B_2}\right) - \left( \frac{y^+}{B_3}\right)\exp\left(-B_4y^+\right)  \right] \, .
    \label{eq:g_formula}
\end{align}
The values of the constants are $B_1 = 9.8$, $B_2 = 3.6$, $B_3 = 5.7$, and $B_4 = 0.27$.
Figure \ref{fig:LoW_and_g_modes} (b) shows a comparison of the $g$ mode from POD at $Re_\tau = 5200$, shown in figure \ref{fig:g_mode_const} (b), with that from \eqref{eq:g_formula}.

\section{Calculation of coefficients used in reconstructions}
\label{App:Calc_coeffs}

To create the reconstructions of the streamwise velocity according to \eqref{eq:EWM} and \eqref{eq:LoW_extension}, we need to calculate the coefficients $c_1$, $c_2$, and $c$.
Thus, we consider a scalar variable $u(x,y,z,t)$ and expand it as follows
\begin{equation}
    u(x,y,z,t) = \sum_{j=1}^N c_j(x,z,t) \varphi_j(y) \, ,
    \label{eq:coeffs_cal_1}
\end{equation}
where $c_j$ are coefficients and $\varphi_j$ are modes.
For non-orthogonal modes $\varphi_j$, such as the $\text{LoW}(y^+)$ and $g(y^+)$, projection of \eqref{eq:coeffs_cal_1} onto the modes $\varphi_j$ leads to a system of linear equations for the coefficients
\begin{equation}
    \sum_{j=1}^N c_j(x,z,t) (\varphi_j , \varphi_k)_\Omega =  (\varphi_j , \varphi_k)_\Omega \, , \qquad k = 1,2,\ldots,N \, ,
\end{equation}
where $(\cdot \, , \, \cdot)_\Omega$ is the inner product defined in \eqref{eq:inner_prod}.
The coefficients can then be determined by solving this linear system of equations.
We note that in practice, where only discrete variables are available, all of the integrals above (appearing in the form of inner products) need to be solved numerically.
In this work, we do this using the trapezoid rule to have consistency with the discretization of the POD eigenvalue problem, see Appendix \ref{App:Num_solv_POD_eig_prob}.

\bibliography{POD_channel}

\end{document}